\documentclass[a4paper,11pt]{article}
\usepackage{jheppub} 

\usepackage[english]{babel}

\usepackage[utf8]{inputenc}
\usepackage{graphicx}
\usepackage{amssymb}
\usepackage{amsmath}
\usepackage[dvipsnames]{xcolor}
\usepackage{hyperref}
\usepackage{multirow}
\usepackage{slashed}
\usepackage{subcaption}
\usepackage{bm}
\usepackage{bbold}
\usepackage{tikz-feynman}

\renewcommand{\vec}[1]{\boldsymbol{#1}}
\newcommand{\be}{\begin{equation}}
\newcommand{\ee}{\end{equation}}
\newcommand{\ba}{\begin{eqnarray}}
\newcommand{\ea}{\end{eqnarray}}
\newcommand{\la}{\label}

\newcommand{\fQ}{f_{\mathcal Q}}

\newcommand{\ahvp}{a_\mu^{\mathrm{HVP}} }

\newcommand{\gpv}[1]{\left[\mathcal{G}(#1)\right]_\Lambda}

\newcommand{\atotvio}{a_\mu^{\text{HVP}, 38}}
\newcommand{\atotvioem}{a_{\mu,em}^{\text{HVP}, 38}}
\newcommand{\atotvioemN}{a_{\mu, em, \text{N451}}^{\text{HVP}, 38}}

\newcommand{\atotvioconn}{a_{\mu,(4)}^{\text{HVP}, 38}}

 \newcommand{\atotvioD}{a_{\mu,D}^{\text{HVP}, 38}}

\begin{document}
\preprint{MITP-26-007, CERN-TH-2026-045}

\author[a]{Dominik~Erb,}
\affiliation[a]{PRISMA$^+$ Cluster of Excellence \& Institut f\"ur Kernphysik,
Johannes Gutenberg-Universit\"at Mainz,
D-55099 Mainz, Germany}
\emailAdd{domerb@uni-mainz.de}

\author[a,b,c]{Harvey~B.~Meyer,}
\affiliation[b]{Helmholtz Institut Mainz, Staudingerweg 18, D-55128 Mainz, Germany}
\affiliation[c]{Theoretical Physics Department, CERN, 1211 Geneva 23, Switzerland}
\emailAdd{h.b.meyer@cern.ch}

\author[d]{ and Konstantin~Ottnad}
\affiliation[d]{Helmholz-Institut f\"ur Strahlen- und Kernphysik and Bethe Center for Theoretical Physics, Universit\"at Bonn, D-53115 Bonn, Germany}
\emailAdd{ottnad@hiskp.uni-bonn.de}

\title{Factorizing the position-space photon propagator in QED corrections to lattice QCD correlators}

\abstract{
Electromagnetic corrections to the $n$-point functions of  lattice QCD can be evaluated using a position-space photon propagator defined in infinite volume. Here we address the computational challenge arising from the volume-squared sum over the endpoints of the photon propagator. We consider a class of integral representations of the photon propagator that lead to a factorization of the two volume-sums, the Fourier representation being one instance thereof. An alternative choice is based on expressing the free scalar propagator as the autoconvolution of the corresponding five-dimensional propagator.
We compare the performance of three different choices in the context of electromagnetic corrections to the hadronic vacuum polarization, on a gauge ensemble of size $48^3\times128$ with a pion mass of 286\;MeV.
As an outlook, we discuss more generally the factorization of sums over internal vertices, taking as an example the hadronic light-by-light contribution to the muon $(g-2)$.
} 
\bigskip

\date{\today}

\maketitle

\section{Introduction \label{sec::Intro}}

The anomalous magnetic moment of the muon serves as a stringent test of the Standard Model (SM) at low energies.
Its direct experimental determination reached an absolute precision of $14.5\times10^{-11}$, corresponding to 124\,ppb~\cite{Muong-2:2025xyk,Muong-2:2023cdq,Muong-2:2024hpx,Muong-2:2021ojo,Muong-2:2021vma,Muong-2:2021ovs,Muong-2:2021xzz,Muong-2:2006rrc}.
The 2025 White Paper (WP25) of the `Muon $g-2$ Theory Initiative' \cite{Aliberti:2025beg} provides the SM prediction with an uncertainty of $62\times10^{-11}$~\cite{Aoyama:2012wk,Volkov:2019phy,Volkov:2024yzc,Aoyama:2024aly,Parker:2018vye,Morel:2020dww,Fan:2022eto,Czarnecki:2002nt,Gnendiger:2013pva,Ludtke:2024ase,Hoferichter:2025yih,RBC:2018dos,Giusti:2019xct,Borsanyi:2020mff,Lehner:2020crt,Wang:2022lkq,Aubin:2022hgm,Ce:2022kxy,ExtendedTwistedMass:2022jpw,RBC:2023pvn,Kuberski:2024bcj,Boccaletti:2024guq,Spiegel:2024dec,RBC:2024fic,Djukanovic:2024cmq,ExtendedTwistedMass:2024nyi,MILC:2024ryz,Bazavov:2024eou,Keshavarzi:2019abf,DiLuzio:2024sps,Kurz:2014wya,Colangelo:2015ama,Masjuan:2017tvw,Colangelo:2017fiz,Hoferichter:2018kwz,Eichmann:2019tjk,Bijnens:2019ghy,Leutgeb:2019gbz,Cappiello:2019hwh,Masjuan:2020jsf,Bijnens:2020xnl,Bijnens:2021jqo,Danilkin:2021icn,Stamen:2022uqh,Leutgeb:2022lqw,Hoferichter:2023tgp,Hoferichter:2024fsj,Estrada:2024cfy,Deineka:2024mzt,Eichmann:2024glq,Bijnens:2024jgh,Hoferichter:2024bae,Holz:2024diw,Cappiello:2025fyf,Colangelo:2014qya,Blum:2019ugy,Chao:2021tvp,Chao:2022xzg,Blum:2023vlm,Fodor:2024jyn}.
This prediction is in agreement with the experimental global average.
However, in order to fully exploit the exquisite experimental precision in ruling out new physics models, it is clearly of high interest to reduce the uncertainty of the SM prediction to a level comparable to that of the experimental global average. 

In the WP25, the hadronic vacuum polarization (HVP) contribution is evaluated using lattice QCD results\footnote{
This is a change from the 2020 WP \cite{Aoyama:2020ynm}, where the dispersive method was used.}.
This O($\alpha^2$) effect is known to a precision of just under one percent, contributing by itself an absolute uncertainty of $61\times 10^{-11}$~\cite{Aliberti:2025beg}.
At the subpercent level, 
not only the purely hadronic vacuum polarization, to be defined in a suitable scheme, but also the corrections 
due to isospin-breaking (IB) effects, both electromagnetic and strong IB, must be taken into account. 
While the size of these corrections appears to be quite small, their uncertainty is the second largest in the overall error budget of the SM prediction. Moreover, the existing evaluations are systematics-dominated~\cite{Blum:2018mom,Borsanyi:2020mff,Djukanovic:2024cmq}. 
With these points in mind, it is clearly of high importance to get a better handle on the corrections caused by IB effects.

Calculating the leading electromagnetic corrections to  lattice QCD correlation functions means including the coupling of quarks to photons. 
This can be done either by including dynamical photons already at the stage of ensemble generation \cite{BMW:2014pzb} (see also the recent study \cite{RCstar:2022yjz}) or by expanding around isosymmetric QCD and perturbatively calculating corrections in the fine structure constant $\alpha$ and in $(m_u-m_d)$, a method first introduced in lattice QCD in 2013 \cite{deDivitiis:2013xla}. 
In the latter class of methods, one can additionally choose to include the photon propagators either in a finite volume, QED$_{\rm L}$, or already in infinite volume, QED$_\infty$. As for the ultraviolet regularization of the photon propagator, it can be chosen to stem from the same lattice used for the QCD degrees of freedom, or from a different regularization. 

In our recent publication~\cite{Erb:2025nxk}, we used the QED$_\infty$ approach with a (doubly) Pauli-Villars regulated photon propagator. 
With this position-space based setup first proposed in~\cite{Biloshytskyi:2022ets}, the photon propagator,   which depends on the relative coordinates of two internal vertices, entangles the volume-wide sums over these vertices.
The most straightforward solution to avoid performing a volume-squared sum is to place a point source on one of the internal vertices~\cite{Erb:2025nxk}. 
This works well, but drastically reduces the sampling of the corresponding vertex.
In this paper we present an alternative approach, where the dependence on the two internal vertices is factorized, which allows the sum over both of them to be carried out at affordable cost. Connected diagrams can thereby be evaluated effectively via the technique of sequential propagators, without introducing additional stochastic noise.

We derive two different implementations of such a factorization and compare them to the method used in our recent publication \cite{Erb:2025nxk}.
Such a task is normally known to be adequately addressed by  methods such as the fast-Fourier transform, however, in the QED$_{\infty}$ approach, the photon propagator results from a superposition of a continuum of momenta, hindering the straightforward application of such methods. The specific factorization formulae we explore, however, are available both for a photon propagator on the infinite lattice or in infinite continuous space -- see appendix \ref{app::Derivation_5D}. 

In general, computing hadronic amplitudes that involve electroweak vertices requires more advanced forms of factorization if the QCD correlation function is to be summed over all but one vertex. Without providing a general solution to this problem, we briefly address, as an important application, the hadronic light-by-light contribution to the muon $(g-2)$ along these lines.

We start by introducing the covariant coordinate-space method as well as the different implementations of the photon kernel in section \ref{sec::Formalism}.
In section \ref{sec::Calculations} we first introduce the ensembles we use throughout this paper.
We then show comparisons between the implementations on ensembles where the gauge links are set to unity. 
The comparison is then extended to the previously introduced QCD ensemble for the fully connected diagrams as well as the relevant disconnected diagrams. 
Section \ref{sec::Adv_Disadv} compares the three implementations by discussing their respective advantages and disadvantages, including their respective computational cost. 
The end of the section shortly compares the results of this work with literature values and gives an outlook on a possible factorization in the hadronic light-by-light calculation.
We conclude in section \ref{sec::Conclusion}.

\section{Formalism \label{sec::Formalism}}
As a starting point and common baseline for comparisons we compute the leading-order isospin-violating part of the HVP contribution to $a_\mu$ as given by the covariant coordinate-space (CCS) method~\cite{Erb:2025nxk},
\begin{align}
    \atotvio(\Lambda)
    &=-\frac{e^2}{2}\int_{z,x,y}  H_{\lambda \sigma}(z) \delta_{\nu \rho}\, [\mathcal{G}(x-y)]_\Lambda \,\big < j^3_\lambda(z) j^{em}_\nu(y) j^{em}_\rho(x) j^8_\sigma (0) \big > + C_T(\Lambda)\, , \label{equ::atotvio}
\end{align}
where $H_{\lambda\sigma}(z)$ is the CCS kernel.
It is not uniquely defined, but has multiple versions, which differ in their shape, but give the same result for $ \ahvp$. In this paper we will focus purely on the traceless ('TL') version of the kernel 
\begin{align}
    \label{eq:tl_kernel}
    H_{\lambda \sigma}^{TL}(z)&=\left(-\delta_{\lambda \sigma}+ 4\frac{z_\lambda z_\sigma}{|z|^2} \right) \mathcal{H}_2(|z|).
\end{align}

Going forward we will drop the 'TL' superscript.
The exact form of the scalar weight function $\mathcal{H}_2$ can be found in~\cite{Meyer:2017hjv} with a rational approximation given in~\cite{Biloshytskyi:2022ets}. 
A version of the scalar weight functions, which correspond to the intermediate window version of the time-momentum representation, can be found in \cite{Chao:2022ycy}.

The second term under the integral, $\delta_{\nu \rho}\, [\mathcal{G}(x-y)]_\Lambda$, is the position space representation of the doubly Pauli-Villars (PV) regulated photon propagator in Feynman gauge, with~\cite{Biloshytskyi:2022ets}
\begin{align}
    \label{equ::pv_photon_prop}
\gpv{x}  =& \frac{1}{4\pi^2 |x|^2} - 2G^{(1)}_{\frac{\Lambda}{\sqrt{2}}}(x) +G^{(1)}_\Lambda(x),\\
G^{(\lambda)}_m(x) =&  \frac{m^\lambda K_\lambda(m |x|)}{(2\pi)^{\lambda+1} |x|^\lambda}.
\label{eq:Gmlda}
\end{align}

It is written with the massive scalar propagator in four dimensions $G^{(1)}_m(x)$, which includes the modified Bessel function of second kind, $K_1(x)$. 
In section~\ref{sec::Factor} we show the different implementations of the photon propagator with their results compared in section~\ref{sec::Calculations}.

The last part in the integral of eq.~\eqref{equ::atotvio} is the vector four-point function. 
It includes two electromagnetic currents at the internal vertices ($em$), an isoscalar current at the origin (3) and an isovector current at the second external vertex (8). 
With the quark-flavour triplet $\vec{\psi}(x) = (u(x) , d(x) , s(x) )^T$,
we define the corresponding currents
\begin{align}
j_\lambda^3 (x) &= \vec{\Bar{\psi}}^T(x) \gamma_\lambda \mathcal{Q}^{(3)} \vec{\psi}(x), \\ 
j_\lambda^8 (x) &= \vec{\Bar{\psi}}^T(x) \gamma_\lambda \mathcal{Q}^{(8)} \vec{\psi}(x), \\ 
j_\lambda^{em} (x) &= \vec{\Bar{\psi}}^T(x) \gamma_\lambda \mathcal{Q}^{(em)} \vec{\psi}(x).    
\end{align}
The respective charge matrices are given by
\begin{align}
\label{equ::charge_matrices}
\mathcal{Q}^{(3)} = \begin{pmatrix}
    1/2 & 0&0 \\
    0& -1/2 &0 \\
    0 & 0 & 0
\end{pmatrix}, \ 
\mathcal{Q}^{(8)} = \begin{pmatrix}
    1/6 & 0&0 \\
    0& 1/6 &0 \\
    0 & 0 & -1/3
\end{pmatrix}, \ 
\mathcal{Q}^{(em)} = \begin{pmatrix}
    2/3 & 0&0 \\
    0& -1/3 &0 \\
    0 & 0 & -1/3
\end{pmatrix} .
\end{align}
The isovector and isoscalar currents provide a decomposition of the electromagnetic current, $j^{em}_\lambda (x) = j^{3}_\lambda (x) + j^{8}_\lambda (x)$.

\begin{figure}[t]
	\centering
	\includegraphics[width=0.6\textwidth]{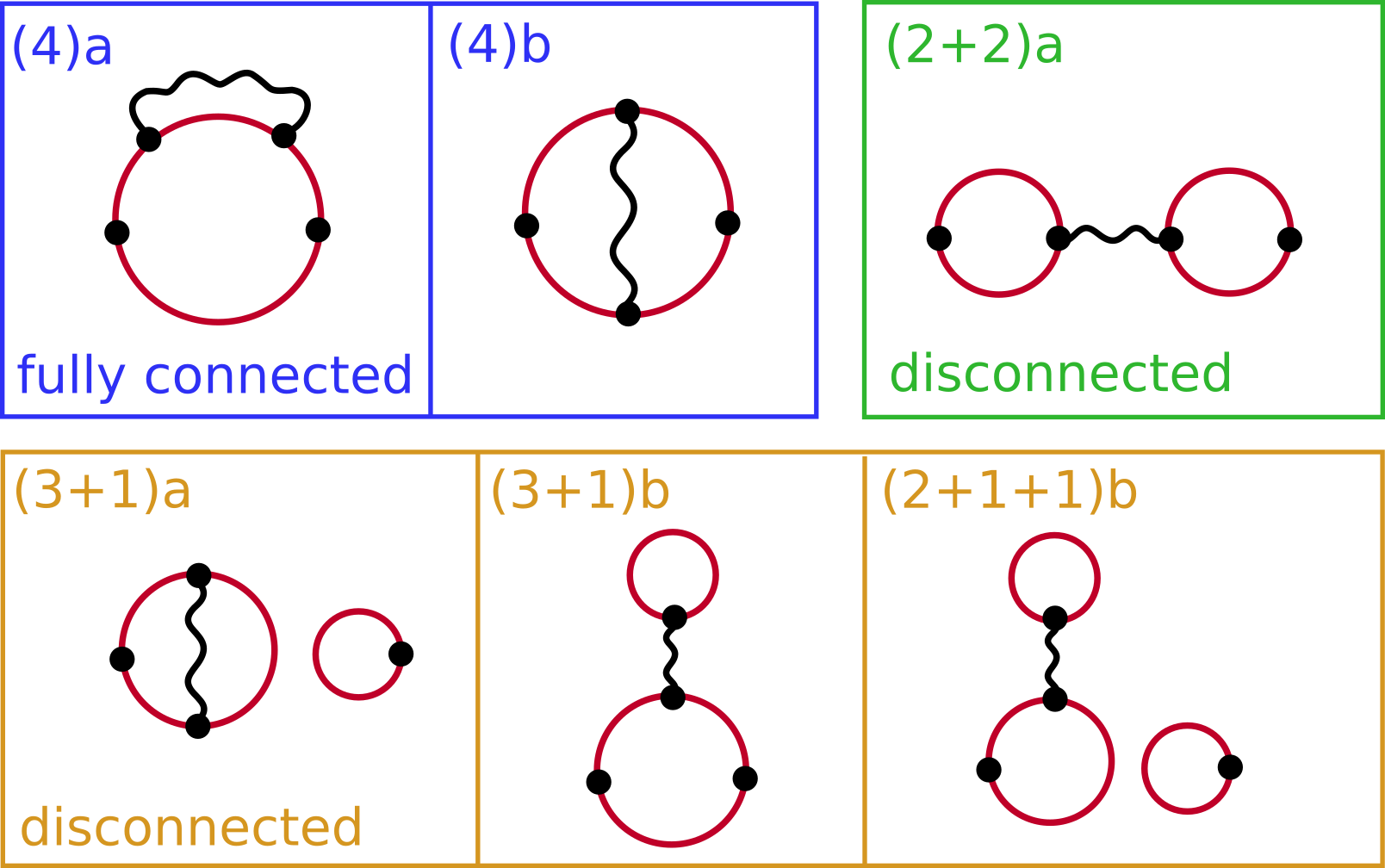}
	\caption{Feynman diagrams depicting all Wick contractions of the four-point function in eq.~\eqref{equ::atotvio}.
    }  
    \label{fig::Formalism_feynman}
\end{figure}

In addition to the integral, eq.~\eqref{equ::atotvio} also includes a counterterm. 
In order to obtain a physical result when taking the limit $\Lambda \rightarrow \infty$, an additional renormalization condition complementary to the scale setting procedure in isospin symmetric QCD  \cite{Bruno:2016plf,Strassberger:2021tsu,RQCD:2022xux} is needed.  
This condition specifies the counterterm and also includes the PV-regulated photon propagator. 
In this work, we purely focus on the first term of eq.~\eqref{equ::atotvio} for a few values of $\Lambda$ and do not consider the counterterm. 

The Wick contractions which contribute to the four point function in eq.~\eqref{equ::atotvio} are shown in figure~\ref{fig::Formalism_feynman}. 
In appendix \ref{app::Charge_factors} we calculate the charge factors of each contraction.
The PV-regularization of the photon propagator is needed for diagrams (4)a, (4)b and (3+1)a, but the other three diagrams (2+2)a, (3+1)b and (2+1+1)b are UV-finite~\cite{Parrino:2025afq}, which makes it possible to use the unregularized version of the photon propagator in these cases, which amounts to taking the limit $\Lambda \rightarrow \infty$ `ahead of time', i.e.\ without a counterterm.

We write the first term of eq.~\eqref{equ::atotvio} for each diagram with only local currents explicitly in terms of propagators:
{\allowdisplaybreaks
\begin{align}
    \atotvioD(\Lambda)&= -\fQ^{D} Z_V^4 \;\frac{e^2} {2}\int_{z,x,y} \ H_{\lambda \sigma}(z) \delta_{\nu \rho} [\mathcal{G}(x-y)]_\Lambda C^{D}_{\rho\nu\lambda\sigma}(x, y, z), \label{equ::atotvioD}\\
    C^{(4)a}_{\rho\nu\lambda\sigma}(x, y, z)&=-4 \, \text{Re} \langle \text{Tr} [ S^l(0, x) \gamma_\rho S^l(x, y) \gamma_\nu S^l(y, z) \gamma_\lambda S^l(z, 0) \gamma_\sigma] \rangle_U, \label{equ::4pt_4a}\\
     C^{(4)b}_{\rho\nu\lambda\sigma}(x, y, z)&=-2 \, \text{Re} \langle \text{Tr} [ S^l(0, y) \gamma_\nu S^l(y, z) \gamma_\lambda S^l(z, x) \gamma_\rho S^l(x, 0) \gamma_\sigma] \rangle_U, \label{equ::4pt_4b}\\
     C^{(2+2)a-ll}_{\rho\nu\lambda\sigma}(x, y, z)&= 2\, \text{Re} \langle \text{Tr} [ S^l(0, x) \gamma_\rho S^l(x, 0) \gamma_\sigma] \times \text{Tr} [ S^l(y, z) \gamma_\lambda S^l(z, y) \gamma_\nu] \rangle_U, \label{equ::4pt2p2all}\\     
     C^{(2+2)a-ls}_{\rho\nu\lambda\sigma}(x, y, z)&= 2\, \text{Re} \langle \text{Tr} [ S^l(0, x) \gamma_\rho S^l(x, 0) \gamma_\sigma] \times \text{Tr} [ S^s(y, z) \gamma_\lambda S^s(z, y) \gamma_\nu] \rangle_U, \label{equ::4pt2p2als}\\     
     C^{(3+1)a}_{\rho\nu\lambda\sigma}(x, y, z)&=2\, \text{Re} \langle \text{Tr} [ S^l(0, y) \gamma_\nu S^l(y, x) \gamma_\rho S^l(x, 0) \gamma_\sigma] \times \text{Tr} [S^\Delta(z,z)\gamma_\lambda ]\rangle_U, \label{equ::4pt3p1a}\\
     C^{(3+1)b}_{\rho\nu\lambda\sigma}(x, y, z)&=4\, \text{Re} \langle \text{Tr} [ S^l(0, y) \gamma_\nu S^l(y, z) \gamma_\lambda  S^l(z, 0) \gamma_\sigma] \times \text{Tr} [S^\Delta(x,x)\gamma_\rho ]\rangle_U, \label{equ::4pt3p1b}\\
     C^{(2+1+1)b}_{\rho\nu\lambda\sigma}(x, y, z)&=-2\, \text{Re} \langle \text{Tr} [ S^l(0, y) \gamma_\nu S^l(y, 0) \gamma_\sigma] \times \label{equ::4pt2p1p1b}\\
     & \qquad \qquad \qquad \qquad \times \text{Tr} [S^\Delta(x,x)\gamma_\rho ] \times \text{Tr} [S^\Delta(z,z)\gamma_\lambda ] \rangle_U, \nonumber\\
     S^\Delta(x,y)&=S^l(x,y)-S^s(x,y). \label{equ::formalism_SDelta}
\end{align}
}
The charge factors $\fQ^D$ are the ones calculated in appendix~\ref{app::Charge_factors}.
We used the notation $S^l(x,y)$ for the light quark propagator going from $y$ to $x$, $S^s(x,y)$ for the strange quark propagator and $S^\Delta(x,y)$ for the difference between the two.
The diagrams have an additional factor of either 2 or 4.
These account for all equivalent Wick contractions contributing to the diagrams. 
There are two effects to consider.
For the diagrams (4)a, (4)b, (3+1)a and (3+1)b the flow of the fermions can be reversed to give a distinct Wick contraction.
The second possibility, which applies to diagram (4)a, (2+2)a, (3+1)b and (2+1+1)b, is the exchange of the $x$- and $y$-vertices, which also gives a distinct Wick contraction. 

Eqs.~\eqref{equ::4pt3p1a} to \eqref{equ::4pt2p1p1b} include  a trace over single disconnected quark loops. 
In the combinations they appear here, their real component is always exactly zero. 
This means for the (3+1) diagrams the real part taken at the end consists purely of the product of the imaginary parts of the two traces. 
On the other hand in the case of the (2+1+1)b diagram only the imaginary parts of the single loops and only the real part of the two-point loop contributes. 

A calculation of the diagrams in the top row of figure~\ref{fig::Formalism_feynman} at the SU(3) flavour symmetric point was done in~\cite{Erb:2025nxk}. Additionally, the (2+2)a diagram was calculated at the physical point with electromagnetic currents instead of the isovector and isoscalar currents in~\cite{Parrino:2025afq}. In section~\ref{sec::Calculations_BareQED} we will first show calculations of the (4)a and (4)b diagrams on ensembles where the gauge links are set to unity. In the section after this we show calculations of the same diagrams, but on a QCD ensemble with a pion mass of 286 MeV. Lastly, section~\ref{sec::Calculations_QCD_Disconnected} will show the bottom row of diagrams on the same ensemble. The (2+2)a diagram will not be discussed here. 

\subsection{Handling the photon propagator \label{sec::Factor}}
In this section we will examine in detail different representations of the photon propagator for our calculations. 
First, we show a more optimized version of the implementation in~\cite{Erb:2025nxk}. 
We then present the general idea of deriving other versions by factorizing the photon propagator. 
This is used to get the two other methods we discuss in this paper. 

\subsubsection{Two-point-source method}

\begin{figure}[t]
		\centering
		\begin{subfigure}{0.24\textwidth}
        \centering
\begin{tikzpicture}[scale=0.5]
    \begin{feynman}
        \vertex at (0, 0) (i) ;
        \vertex at (1,0) [dot](a){};
        \vertex at (1.293,0.707) [empty dot](b){};
        \vertex at (2.707,0.707) [ dot](c){};
        \vertex at (3,0) [empty dot](d){};
        \vertex at (2,-1) (e);
        \vertex at (4,0) (f) ;
        \diagram*{
            (i) -- [photon, thick] (a)  ,
            (a) -- [out=90, in= 225, looseness=0.8, very thick, red] (b) -- [out=45, in=135, very thick, Orange] (c) -- [ out=-45, in=90, looseness=0.8, very thick, Green] (d) --[quarter left, very thick, blue] (e) -- [quarter left, very thick, blue] (a) ,
            (d) -- [photon, thick] (f),
            (b) -- [half left, photon, very thick] (c);
        };
    \end{feynman}
\end{tikzpicture}
			\caption{(4)a}
		\end{subfigure}
		\begin{subfigure}{0.24\textwidth}
        \centering
\begin{tikzpicture}[scale=0.5]
    \begin{feynman}
        \vertex at (0, 0) (i) ;
        \vertex at (1, 0) [dot](a){};
        \vertex at (2,1) [dot](b) {};
        \vertex at (2, -1) [empty dot] (c){};
        \vertex at (3, 0) [empty dot](d){};
        \vertex at (4,0) (f) ;
        \diagram*{
            (i) -- [photon, thick] (a) ,
            (a) -- [quarter left, red, very thick] (b)  -- [quarter left, Green, very thick] (d) -- [quarter left, blue, very thick] (c) --[quarter left, blue, very thick] (a),
            (d) -- [photon, thick] (f),
            (b) -- [photon, blue, very thick] (c);
        };
    \end{feynman}
\end{tikzpicture}
			\caption{(4)b}
		\end{subfigure}
        
		\begin{subfigure}{0.24\textwidth}
        \centering
\begin{tikzpicture}[scale=0.5]
    \begin{feynman}
        \vertex at (0, 0) (i) ;
        \vertex at (1, 0) [dot](a){};
        \vertex at (1.75, 0.75) [dot](b) {};
        \vertex at (1.75, -0.75) [empty dot] (c) {};
        \vertex at (4, 0) [empty dot](d){};
        \vertex at (3,0) (d1);
        \vertex at (5,0) (f) ;
        \vertex at (2.5, 0) (g);
        \diagram*{
            (i) -- [photon, thick] (a) ,
            (a) -- [quarter left, red, very thick] (b)  -- [quarter left, Green, very thick] (g) -- [quarter left, Green, very thick] (c) --[quarter left, blue, very thick] (a),
            (d) -- [photon, thick] (f),
            (b) -- [photon, very thick] (c),
            (d1) -- [half left, orange, very thick] (d),
            (d) -- [half left, orange, very thick] (d1);
        };
    \end{feynman}
\end{tikzpicture}
			\caption{(3+1)a}
		\end{subfigure}        
		\begin{subfigure}{0.24\textwidth}
        \centering
\begin{tikzpicture}[scale=0.5]
    \begin{feynman}
        \vertex at (0, 0) (i) ;
        \vertex at (1, 0) [dot](a){};
        \vertex at (1.75, 0.75) [dot](b) {};
        \vertex at (1.75, -0.75) (c);
        \vertex at (2.5, 0) [empty dot](d){};
        \vertex at (3.5,0) (f) ;
        \vertex at (1.75, 2) [empty dot] (d1) {};
        \vertex at (1.75, 3) (d2);
        \diagram*{
            (i) -- [photon, thick] (a) ,
            (a) -- [quarter left, red, very thick] (b)  -- [quarter left, Green, very thick] (d) -- [quarter left, blue, very thick] (c) --[quarter left, blue, very thick] (a),
            (d) -- [photon, thick] (f),
            (b) -- [photon, very thick] (d1);
            (d1) -- [half left, orange, very thick] (d2),
            (d2) -- [half left, orange, very thick] (d1);
        };
    \end{feynman}
\end{tikzpicture}
			\caption{(3+1)b}
		\end{subfigure}
		\begin{subfigure}{0.24\textwidth}
        \centering
\begin{tikzpicture}[scale=0.5]
    \begin{feynman}
        \vertex at (0, 0) (i) ;
        \vertex at (1, 0) [dot](a){};
        \vertex at (1.75, 0.75) [dot](b) {};
        \vertex at (1.75, -0.75) (c);
        \vertex at (2.5, 0) (d);
        \vertex at (5,0) (f) ;
        \vertex at (1.75, 2) [empty dot] (d1) {};
        \vertex at (1.75, 3) (d2);
        \vertex at (3,0) (g1);
        \vertex at (4,0) [empty dot] (g2){};
        \diagram*{
            (i) -- [photon, thick] (a) ,
            (a) -- [quarter left, red, very thick] (b)  -- [quarter left, blue, very thick] (d) -- [quarter left, blue, very thick] (c) --[quarter left, blue, very thick] (a),
            (g2) -- [photon, thick] (f),
            (b) -- [photon, very thick] (d1);
            (d1) -- [half left, orange, very thick] (d2),
            (d2) -- [half left, orange, very thick] (d1);
            (g1) -- [half left, Green, very thick] (g2),
            (g2) -- [half left, Green, very thick] (g1);
        };
    \end{feynman}
\end{tikzpicture}
			\caption{(2+1+1)b}
            \label{fig::Feynman_diagrams_2PS_2p1p1b}        
		\end{subfigure}
\caption{The diagrams discussed in this work as one would calculate them with the 2PS method. The filled black dots represent the vertices from which the one-to-all propagators start, while the unfilled dots represent the vertices which get summed over. Each color is a separate propagator. Only diagram (4)b includes a sequential propagator, which goes over one of the internal vertices and includes the photon propagator.}
\label{fig::Feynman_diagrams_2PS}        
\end{figure}
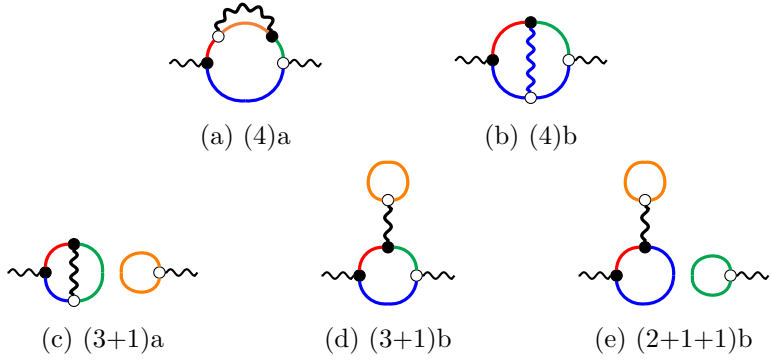

As the name suggests, in the two-point-source method (2PS) we use two point sources to calculate the diagrams, one at the origin and one at the internal vertex $y$. 
It starts out by using rotational invariance to reduce the integral over the $y$-vertex in eq.~\eqref{equ::atotvio} to a one-dimensional integral
\begin{align}
    \atotvioem(\Lambda) =-\pi^2 {e^2}  \int_0^\infty  d|y|  \,  |y|^3 a^8\sum_{z,x}& H_{\lambda \sigma}(z) \,\delta_{\nu \rho}\,[\mathcal{G}(x-y)]_\Lambda \times \\
    &\times \big < j^3_\lambda(z) j^{em}_\nu(y) j^{em}_\rho(x) j^8_\sigma (0) \big >.\nonumber
\end{align}
We also wrote the integrals over the remaining two vertices explicitly as sums over the lattice positions. 
We now want to use multiple combinations of origin and $y$-vertex to reconstruct the integral over $|y|$. 
First, a starting point, which sits on the center time-slice\footnote{For an ensemble with (anti)-periodic boundary conditions in time, the time-slice can be chosen arbitrarily.}, but has for each configuration a randomly chosen spatial position, is picked.
Next, a one-to-all propagator starting at this point is calculated and stored.
This step gets repeated for all points which sit along the diagonal which goes through the starting point and has the $(x,y,z,t)=(1,1,1,0)$ direction\footnote{The setup would work for each diagonal of the form $(\pm 1, \pm 1, \pm 1, 0)$. 
}. 
Since the spatial lengths of our ensembles are always equal, the diagonal wraps around on itself. 
With this setup we can pick each point on the diagonal as the origin and all other points as the $y$-vertex, which increases our statistics. In order to save some memory and computing time, only every second point along the diagonal is calculated. 

Figure~\ref{fig::Feynman_diagrams_2PS} gives an overview on how the propagators are combined to calculate each diagram. 
The filled dots are the vertices from which the one-to-all propagators start, while the unfilled dots are the vertices which get summed over. 
For all diagrams except (4)b the so far mentioned one-to-all propagators and their all-to-one counterparts are sufficient. 
But for this last diagram we also need to calculate a sequential propagator, which starts at the origin, goes over the $x$-vertex, the second internal one, and ends at the $z$-vertex. 
During the calculation of the sequential propagator the photon propagator needs to be incorporated, which makes it impossible to calculate multiple PV-masses at the same time.

Since the calculation of the (4)b diagram is by far the biggest time sink, we apply two adjustments.
First, instead of using each point on the diagonal as an origin, we only use four of them. 
Additionally, we make use of the truncated solver method \cite{Bali:2009hu} when calculating the sequential propagator.  
For our usual solves we use a residue of $1\times 10^{-10}$ as the stopping condition. 
But for the truncated solves this gets relaxed to $1\times 10^{-4}$. 
The resulting speed-up is around a factor of three.
To account for potential biases we also do an exact solve for the first origin point and use this for a bias correction.

The methodology described here works exceptionally well for the (4)a diagram, but the additional cost for the (4)b diagram is very high. 
Reducing this cost is one of the reasons why we developed the alternative methods in the next sections.

\subsubsection{Factorizing the photon propagator}

\begin{figure}[t]
		\centering
		\begin{subfigure}{0.24\textwidth}
        \centering
\begin{tikzpicture}[scale=0.5]
    \begin{feynman}
        \vertex at (0, 0) (i) ;
        \vertex at (1,0) [dot](a){};
        \vertex at (1.293,0.707) [empty dot](b){};
        \vertex at (2.707,0.707) [empty dot](c){};
        \vertex at (3,0) [empty dot](d){};
        \vertex at (2,-1) (e);
        \vertex at (4,0) (f) ;
        \diagram*{
            (i) -- [photon, thick] (a)  ,
            (a) -- [out=90, in= 225, looseness=0.8, very thick, red] (b) -- [out=45, in=135, very thick, red] (c) -- [ out=-45, in=90, looseness=0.8, very thick, red] (d) --[quarter left, very thick, blue] (e) -- [quarter left, very thick, blue] (a) ,
            (d) -- [photon, thick] (f),
            (b) -- [half left, photon, red, very thick] (c);
        };
    \end{feynman}
\end{tikzpicture}
			\caption{(4)a}
		\end{subfigure}
		\begin{subfigure}{0.24\textwidth}
        \centering
\begin{tikzpicture}[scale=0.5]
    \begin{feynman}
        \vertex at (0, 0) (i) ;
        \vertex at (1, 0) [dot](a){};
        \vertex at (2,1) [empty dot](b) {};
        \vertex at (2, -1) [empty dot] (c){};
        \vertex at (3, 0) [empty dot](d){};
        \vertex at (4,0) (f) ;
        \vertex at (2,0) (ph);
        \diagram*{
            (i) -- [photon, thick] (a) ,
            (a) -- [quarter left, red, very thick] (b)  -- [quarter left, red, very thick] (d) -- [quarter left, blue, very thick] (c) --[quarter left, blue, very thick] (a),
            (d) -- [photon, thick] (f),
            (c) -- [photon, blue, very thick] (ph);
            (b) -- [photon, red, very thick] (ph);
        };
    \end{feynman}
\end{tikzpicture}
			\caption{(4)b}
		\end{subfigure}
        
		\begin{subfigure}{0.24\textwidth}
        \centering
\begin{tikzpicture}[scale=0.5]
    \begin{feynman}
        \vertex at (0, 0) (i) ;
        \vertex at (1, 0) [dot](a){};
        \vertex at (1.75, 0.75) [empty dot](b) {};
        \vertex at (1.75, -0.75) [empty dot] (c) {};
        \vertex at (4, 0) [empty dot](d){};
        \vertex at (3,0) (d1);
        \vertex at (5,0) (f) ;
        \vertex at (2.5, 0) (g);
        \vertex at (1.75, 0) (ph);
        \diagram*{
            (i) -- [photon, thick] (a) ,
            (a) -- [quarter left, red, very thick] (b)  -- [quarter left, red, very thick] (g) -- [quarter left, red, very thick] (c) --[quarter left, blue, very thick] (a),
            (d) -- [photon, thick] (f),
            (b) -- [photon, red, very thick] (ph),
            (ph) -- [photon, blue, very thick] (c),
            (d1) -- [half left, orange, very thick] (d),
            (d) -- [half left, orange, very thick] (d1);
        };
    \end{feynman}
\end{tikzpicture}
			\caption{(3+1)a}
		\end{subfigure}        
		\begin{subfigure}{0.24\textwidth}
        \centering
\begin{tikzpicture}[scale=0.5]
    \begin{feynman}
        \vertex at (0, 0) (i) ;
        \vertex at (1, 0) [dot](a){};
        \vertex at (1.75, 0.75) [empty dot](b) {};
        \vertex at (1.75, -0.75) (c);
        \vertex at (2.5, 0) [empty dot](d){};
        \vertex at (3.5,0) (f) ;
        \vertex at (1.75, 2) [empty dot] (d1) {};
        \vertex at (1.75, 3) (d2);
        \vertex at (1.75, 1.375) (ph);
        \diagram*{
            (i) -- [photon, thick] (a) ,
            (a) -- [quarter left, red, very thick] (b)  -- [quarter left, red, very thick] (d) -- [quarter left, blue, very thick] (c) --[quarter left, blue, very thick] (a),
            (d) -- [photon, thick] (f),
            (b) -- [photon, very thick, red] (ph);
            (d1) -- [photon, very thick, orange] (ph);
            (d1) -- [half left, orange, very thick] (d2),
            (d2) -- [half left, orange, very thick] (d1);
        };
    \end{feynman}
\end{tikzpicture}
			\caption{(3+1)b}
		\end{subfigure}
		\begin{subfigure}{0.24\textwidth}
        \centering
\begin{tikzpicture}[scale=0.5]
    \begin{feynman}
        \vertex at (0, 0) (i) ;
        \vertex at (1, 0) [dot](a){};
        \vertex at (1.75, 0.75) [empty dot](b) {};
        \vertex at (1.75, -0.75) (c);
        \vertex at (2.5, 0) (d);
        \vertex at (5,0) (f) ;
        \vertex at (1.75, 2) [empty dot] (d1) {};
        \vertex at (1.75, 3) (d2);
        \vertex at (3,0) (g1);
        \vertex at (4,0) [empty dot] (g2){};
        \vertex at (1.75, 1.375) (ph);
        \diagram*{
            (i) -- [photon, thick] (a) ,
            (a) -- [quarter left, red, very thick] (b)  -- [quarter left, blue, very thick] (d) -- [quarter left, blue, very thick] (c) --[quarter left, blue, very thick] (a),
            (g2) -- [photon, thick] (f),
            (b) -- [photon, very thick, red] (ph);
            (d1) -- [photon, very thick, orange] (ph);
            (d1) -- [half left, orange, very thick] (d2),
            (d2) -- [half left, orange, very thick] (d1);
            (g1) -- [half left, Green, very thick] (g2),
            (g2) -- [half left, Green, very thick] (g1);
        };
    \end{feynman}
\end{tikzpicture}
			\caption{(2+1+1)b}
		\end{subfigure}
\caption{The diagrams discussed in this work as one would calculate them with the factorization methods. The filled black dots represent the point source from which the one-to-all propagators are calculated, while the unfilled dots represent the vertices which get summed over. Each color is a separate propagator. If a propagator goes over an unfilled dot it is a (double) sequential propagator over that vertex. }
\label{fig::Feynman_diagrams_Factor}        

\end{figure}
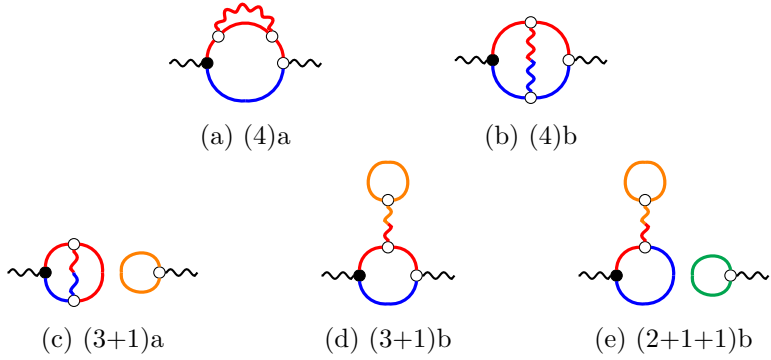

The previous methodology set one of the internal vertices on a single point in order to avoid the volume-square sum. It is needed, because the photon propagator depends on the difference of the two internal vertices. 
Alternatively, one could also try to disentangle the two volume sums. Generally, this could be written as follows
\begin{align}
    \gpv{x-y}=\int d\xi \ \phi_\Lambda(\xi) h^*_{\xi,\Lambda}(x) g_{\xi,\Lambda}(y).
\end{align}
Inserting this expression into Eq.~\eqref{equ::atotvio}, one obtains
\begin{align}
    \atotvioem(\Lambda)
    =-\frac{e^2}{2} \int d\xi \ \phi_\Lambda(\xi) \, a^{12}\sum_{z,x,y}  &H_{\lambda \sigma}(z) \delta_{\nu \rho}\, h^*_{\xi, \Lambda}(x) g_{\xi, \Lambda}(y) \times\label{equ::atotvio_gamma}\\
    &\times \big < j^3_\lambda(z) j^{em}_\nu(y) j^{em}_\rho(x) j^8_\sigma (0) \big >. \nonumber
\end{align}
Such a setup now allows for the summation over both internal vertices as well as one of the external ones, at the cost of having an additional integral over the parameter $\xi$. 
As compared to the 2PS method, the computational strategy changes:
The origin is chosen in the same way as previously described for the starting point and a one-to-all propagator starting from this point is calculated. 
For the (4)b diagram, a sequential propagator starting at the origin, going over the $x$-vertex and ending at the $z$-vertex is calculated, as well as one going over the $y$-vertex. These include $h^*_{\xi, \Lambda}(x)$ and $ g_{\xi, \Lambda}(y)$ respectively. 
The disconnected diagrams can be calculated by reusing the propagators from the (4)b diagram. 
The (4)a diagram is now the one needing more computing time. To calculate it, a double sequential propagator starting at the origin, going over the $x$- and $y$-vertices and ending at the $z$-vertex is necessary. Both $h^*_{\xi, \Lambda}(x)$ and $ g_{\xi, \Lambda}(y)$ are included in this calculation. 
As in the previous section, an overview on how the calculated propagators are put together for the different diagrams is given in figure~\ref{fig::Feynman_diagrams_Factor}.

The computation has to be repeated for each choice of the PV-mass and for each value of $\xi$ needed for computing the final integral. 
This makes it not obvious if a method following this new approach is computationally more efficient than the 2PS method, especially since the latter allows for the optimization of using the diagonal. 
But the factorization method has also two potential optimizations we have not discussed so far.
If $h_{\xi, \Lambda} =g_{\xi, \Lambda}$, only one sequential propagator needs to be calculated for the (4)b and the disconnected diagrams, because the second one can be obtained by inverting the first one. 
Additionally, if $h$ and $g$ are independent of $\Lambda$, multiple PV-masses can be evaluated at the same time, saving potentially a lot of computing time. 

Now, we introduce two different factorizations of the photon propagator which include one or both of the discussed optimizations. 

\subsubsection{Fourier method}
Both optimizations are included in our first variation. We use the following setup
\begin{align}
    \xi:=&\ k\in \mathbb{R} ^4\\
    \phi_\Lambda(k)=&\ \frac{1}{k^2}-\frac{2}{k^2+\Lambda^2/2}+\frac{1}{k^2+\Lambda^2}\\
    h_{k, \Lambda}(x) =&\ g_{k, \Lambda}(x)=e^{ikx}
\end{align}

We call it the Fourier method as we revert the photon propagator back to its momentum-space representation. 
Again, we use rotational invariance to reduce the four dimensional integral over $k$ to a 1D integral over $|k|$.
$k$ is chosen along the axis $(1,1,1,1)$. For large $k$ the step size is of the physical lattice momentum, i.e. $2\pi/L$ and $2\pi/T$ respectively, with the lattice space extend $L$ and time extend $T$. 
The integrand is peaked within the first few of these steps. 
To still cover it appropriately we also compute unphysical momenta in between these steps. 

\subsubsection{5D propagator method\label{sec:5dpropmeth}}

The second variation separately factorizes each of the three scalar propagators contributing to the regularized photon propagator,
\begin{align}
    G^{(1)}_m(x-y)=&\int d\xi \ \phi_m(\xi) h^*_{\xi,m}(x) g_{\xi,m}(y)\\
    \xi:=&\ w\in \mathbb{R}^4\\
    \phi_m(w)=&\ 4\\
    h_{w,m}(x)=&\ g_{w,m}(x)=G^{(3/2)}_m(w-x)\\
    G^{(3/2)}_m(w-x)=&\ \frac{m|w-x| +1}{8\pi^2 |w-x|^3}\,e^{-m |w-x|} .
\end{align}
Thus each four-dimensional scalar propagator, $G_{\frac{\Lambda}{\sqrt{2}}}^{(1)}(x-y)$, $G_{{\Lambda}}^{(1)}(x-y)$ and the massless one, is split into the product of two five dimensional scalar propagators, hence the name 5D propagator method.
A derivation of this factorization is provided in appendix \ref{app::Derivation_5D}.

The point $w$ can be interpreted as the 'mid-point' of the photon propagator. As such it is not restricted to lattice sites and we can choose arbitrary positions for it. 
As with the previous two methods, we use rotational invariance to reduce the four-dimensional integral over $w$ to a 1D integral over $|w|$. 
We choose the starting point to be $(1/2, 1/2, 1/2, 1/2)$ in lattice units, because this avoids any numerical problems that might arise for $w=x$ or $w=y$. 
We let $w$ vary in integer steps along the $(1,1,1,1)$ diagonal, but, as for the 2PS method, we only compute every second point. 

Since the three terms of the photon propagator are factorized separately, we cannot calculate them together like it is done in the 2PS method. 
That means we essentially have to repeat the calculation three times. 
But for additional PV-masses this is reduced to either two additional calculations, as the $m=0$ term is already computed, or even only one additional calculation, if the set of PV masses is suitably chosen.

\section{Lattice calculations \label{sec::Calculations}}
In this paper we will calculate the leading-order isospin-violating part of hadronic vacuum polarization in the muon $(g-2)$, as discussed in~\cite{Erb:2025nxk}, in order to compare the three different versions of the photon propagator. Similar to the strategy pursued there, we will first perform the calculations of the connected contributions on ensembles without the strong interaction, i.e.\ where the gauge field variables $U\in {\rm SU}(3)_c$ are set to unity on all lattice links. 
Owing to the use of the Pauli-Villars regulated photon propagator (as opposed to the regularization by the finite lattice spacing), we can directly compare the lattice results to continuum calculations (see appendix~B of \cite{Erb:2025nxk}). This allows us to accurately compare the three versions of the photon propagator. Following this, we will do the same comparison on a QCD ensemble with a pion mass of $m_\pi=$ 286(3) MeV. We finish this section by also comparing the disconnected contributions.

\subsection{Lattice setup \label{sec::LatSet}}
Let us first discuss the ensembles we use throughout this paper.
For the crosscheck with continuum calculations we employ eight $L^4$ ensembles, where the gauge field variables are set to unity. They have different lattice spacings, but a constant volume of $m_\mu L=7.2$.
The lattice spacings are provided in table~\ref{tab::Conn_QED_Lattice_Parameters}. 
In order to correct for finite size effects we also use one additional ensemble with $L=48$ and $m_\mu L=14.4$. 
It has the same parameters as the $L=24$ ensemble but with twice the lattice lengths.

We perform our lattice QCD calculations on N451, one of the gauge ensembles generated by the CLS initiative \cite{Bruno:2014jqa, Bali:2016umi}, with dynamical $O(a)$ improved Wilson-Clover up, down and strange quarks and the tree-level $O(a^2)$ improved Lüscher-Weisz gauge action. It has periodic boundary conditions in time and a pion mass of $m_\pi = 286(3)$ MeV. More details are shown in table~\ref{tab::LatSet_Properties}. 

\begin{table}[t]
	\centering
	\begin{tabular}{c c c c c c c c c}
		\hline
		
		$L$ & 24  & 32 & 40 & 48 & 56 & 64 & 72 & 80 \\ \hline
		$am_\mu$ & 0.3 & 0.225 & 0.18 & 0.15 & 0.13 & 0.1125 & 0.1 & 0.09\\ \hline  
	\end{tabular}
	\caption{Parameters of the gluon-less ensembles. For each lattice the time extent is the same as $L$, i.e. they have a volume of $L^4$.}
	\label{tab::Conn_QED_Lattice_Parameters}
\end{table}

\begin{table}[t]
\centering
    \begin{tabular}{ c c c c c c c c c c}
    \hline
        Id & $\beta$ & $(\frac{L}{a})^3 \times \frac{T}{a}$ & a [fm] & $m_\pi$ [MeV]& $m_K$ [MeV]  & $m_\pi L$ & L[fm] & $\hat{Z}_V$\\ \hline
        N451 & 3.46 & $48^3 \times 128$ & 0.07634 & 286(3) & 461(5)  & 5.3 & 3.7 & 0.72645 \\ \hline
    \end{tabular}
    \caption{Parameters of N451. The lattice spacing in physical units was extracted from~\cite{Bruno:2016plf}. The pion and kaon mass values are taken from~\cite{Ce:2022kxy}. The value of the full renormalization factor $\hat{Z}_V$ for the local current is from Ref.~\cite{Gerardin_2019}. N451 has periodic boundary conditions in time.}
    \label{tab::LatSet_Properties}

\end{table}

In order to improve future continuum extrapolations, we use two different discretizations for the vector current, the local $(l)$ and the point-split or conserved $(c)$ discretizations 
\begin{align}
    j^l_\nu(z)&=\Bar{\psi}(z)\gamma_\nu \mathcal{Q}\psi(z), \\ 
    j^a_\nu(z)&= \frac{1}{2} \left[\bar{\psi}(z+a\hat{\nu}) (\gamma_\nu +1) U^\dagger_{\nu}( z) \, \mathcal{Q} \psi(z) + \bar{\psi}(z) (\gamma_\nu-1) U_{\nu}(z) \, \mathcal{Q} \psi(z+a\hat{\nu})\right],  \\
    j^c_\nu(z)&=\frac{1}{2}\big( j^a_\nu(z)+ j^a_\nu(z-a\hat{\nu})\big).
\end{align}

$U_{\nu}(z)$ is the gauge link in the direction $\nu$ associated with site $z$. 
No further improvements of the vector currents are implemented in this paper. 
To denote the different combinations of local and conserved currents we adopt the notation `XdYdZd' for the different discretizations. 
The `d' is replaced by either an `l' for a local current at the corresponding vertex or a `c' for a conserved current. 
The current at the origin is always a local one.

While the conserved vector currents do not need to be renormalized, a renormalization factor $\hat{Z}_V$ is required for the local currents. 
The renormalized local vector current is obtained from $j^{l,\text{ren}}_\nu(z)= \hat{Z}_V j^l_\nu(z)$. 
We extracted $\hat{Z}_V$ from Ref.\ \cite{Gerardin_2019}; the value is noted in table~\ref{tab::LatSet_Properties}.

\subsection{Bare QED two-loop vacuum polarization contribution at fixed $\Lambda$ from the lattice \label{sec::Calculations_BareQED}}
For this crosscheck we set the lepton mass $m_\ell$ appearing in the loop to be equal to the muon mass, $m_\ell=m_\mu$. 
We use a Pauli-Villars mass of $\Lambda=3\, m_\mu$, which is the same mass used for the `gluon-less' crosscheck performed in \cite{Erb:2025nxk}.

\begin{figure}[t]
    \centering
    \begin{subfigure}{0.32\textwidth}
        \includegraphics[width=\textwidth]{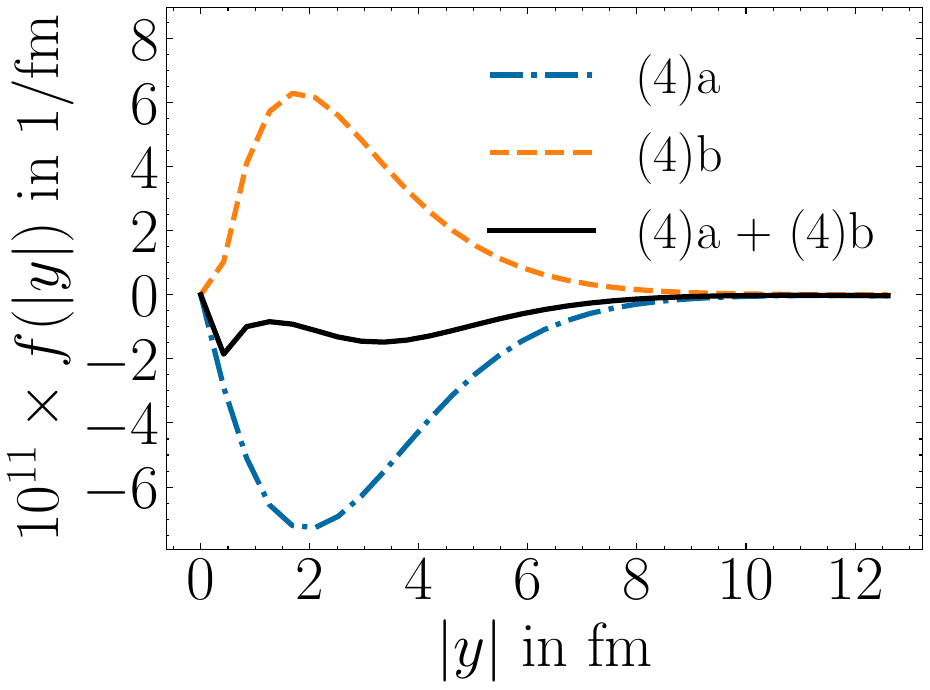}
        \caption{2PS method }
    \label{fig::Conn_QED_Integrands_2PS}
    \end{subfigure}
    \begin{subfigure}{0.32\textwidth}
        \includegraphics[width=\textwidth]{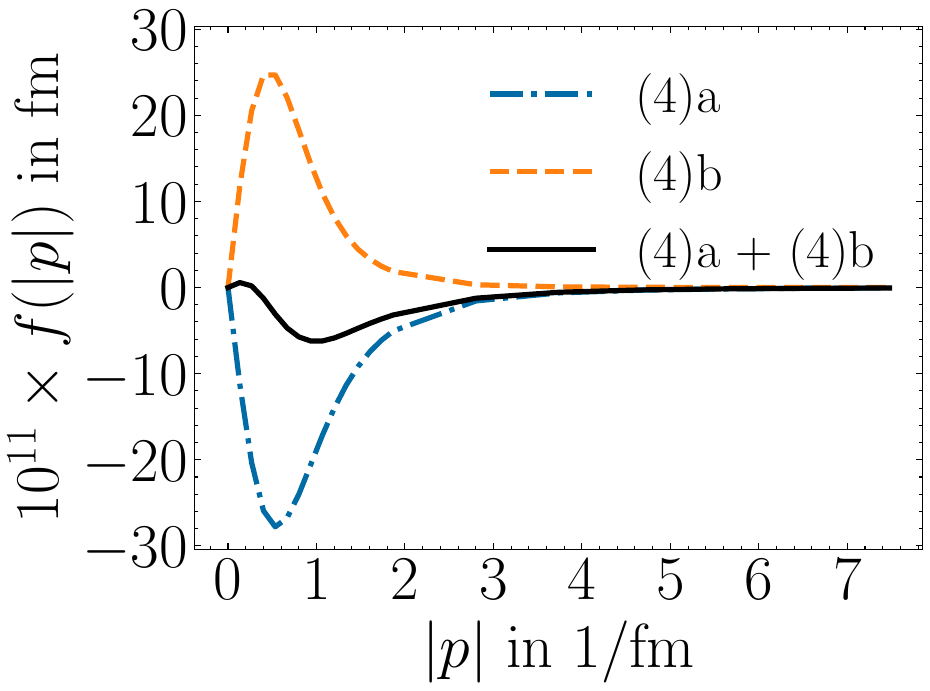}
        \caption{Fourier method }
    \label{fig::Conn_QED_Integrands_Fourier}
    \end{subfigure}
    \begin{subfigure}{0.32\textwidth}
        \includegraphics[width=\textwidth]{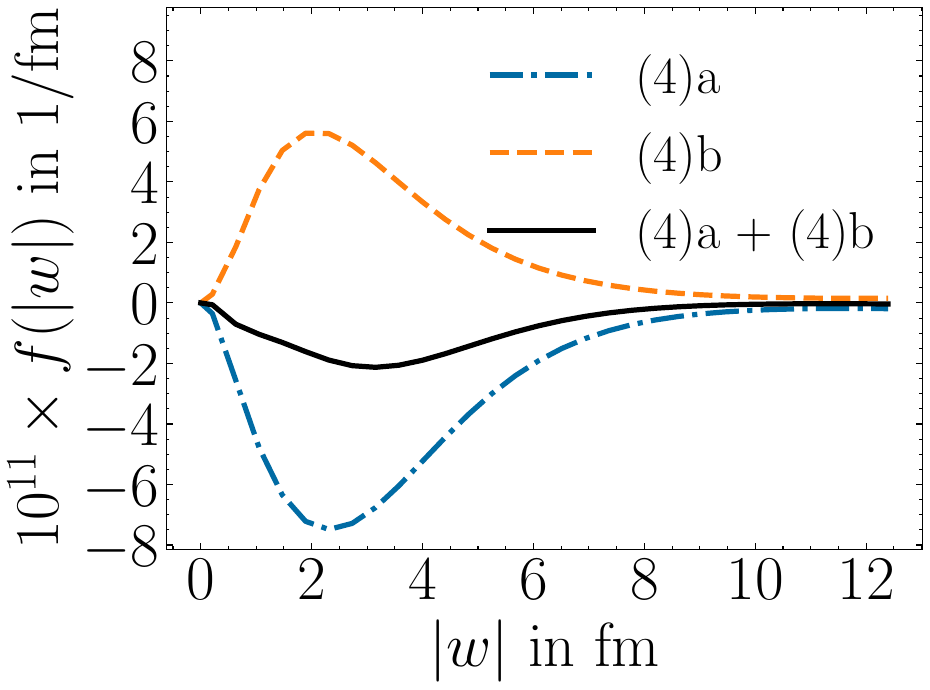}
        \caption{5D propagator method }
    \label{fig::Conn_QED_Integrands_5D}
    \end{subfigure}
    \caption{Final integrands for the different methods with $\Lambda = 3\, m_\mu$ and the XcYlZc discretization on the gluon-less ensemble with $L=64$. The lepton in the loop has mass $m_\ell=m_\mu$. We show the (4)a diagram (blue) and the (4)b diagram (orange) separately as well as their sum (black). }    
    \label{fig::Conn_QED_Integrands}
    
\end{figure}

An example of the final integrand of each method can be seen in figure~\ref{fig::Conn_QED_Integrands}. The ensemble is the one with $L=64$ and the discretization is XcYlZc.
It is important to keep in mind that the variables on the horizontal axis are different in the three panels. 
For the left panel,
it is the absolute value of the difference between the origin and one of the internal vertices, 
for the middle panel
it is the absolute value of the photon momentum, and 
for the right panel
it is the absolute value of the difference between the origin and the `mid-point' of the photon propagator.

\begin{figure}[t]
    \centering
    \begin{subfigure}{\textwidth}
        \includegraphics[width=\textwidth]{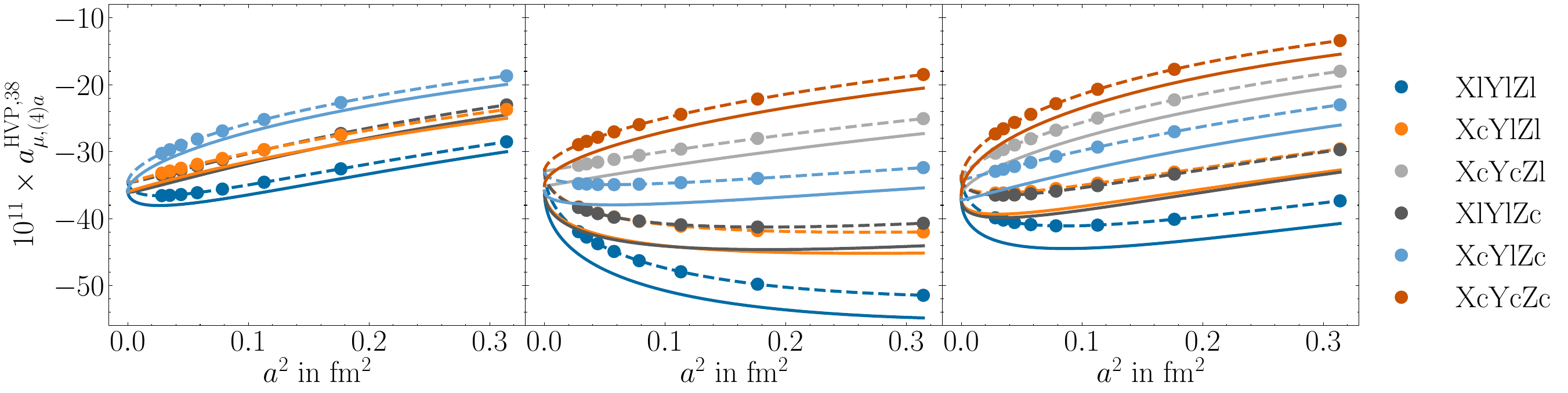}
    \end{subfigure}    

        \hspace{-1cm}
    \begin{subfigure}{0.19\textwidth}
    \caption{}
    \end{subfigure}
    \begin{subfigure}{0.32\textwidth}
    \caption{}
    \end{subfigure}
    \begin{subfigure}{0.19\textwidth}
    \caption{}
    \end{subfigure}
    \vspace{0.3cm}
    
    \begin{subfigure}{\textwidth}
        \includegraphics[width=\textwidth]{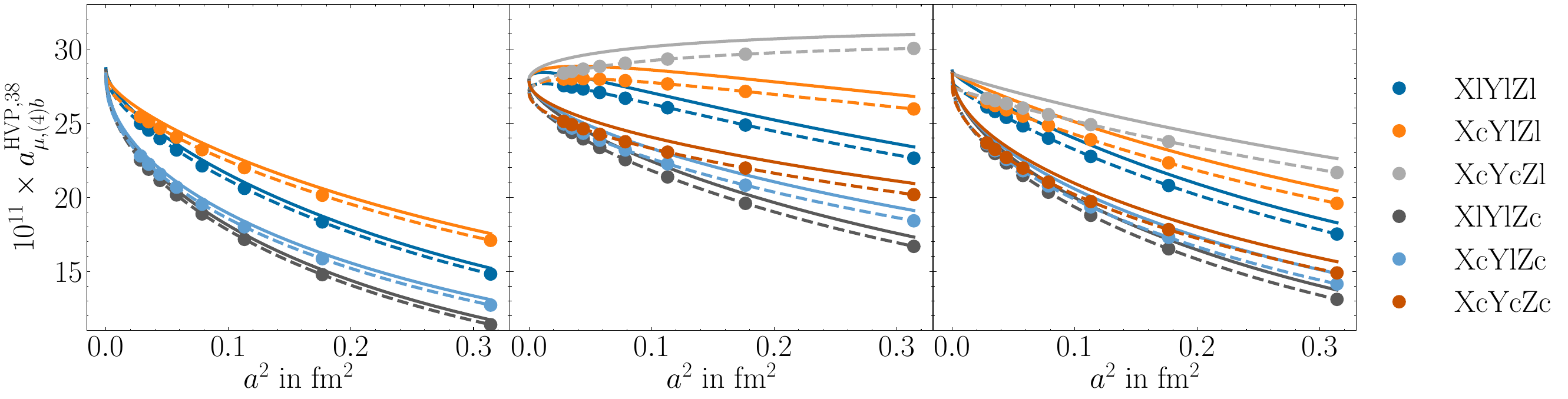}
    \end{subfigure}
    
    \hspace{-1cm}
    \begin{subfigure}{0.19\textwidth}
    \caption{}
    \end{subfigure}
    \begin{subfigure}{0.32\textwidth}
    \caption{}
    \end{subfigure}
    \begin{subfigure}{0.19\textwidth}
    \caption{}
    \end{subfigure}
    \vspace{0.3cm}
    
    \begin{subfigure}{\textwidth}
        \includegraphics[width=\textwidth]{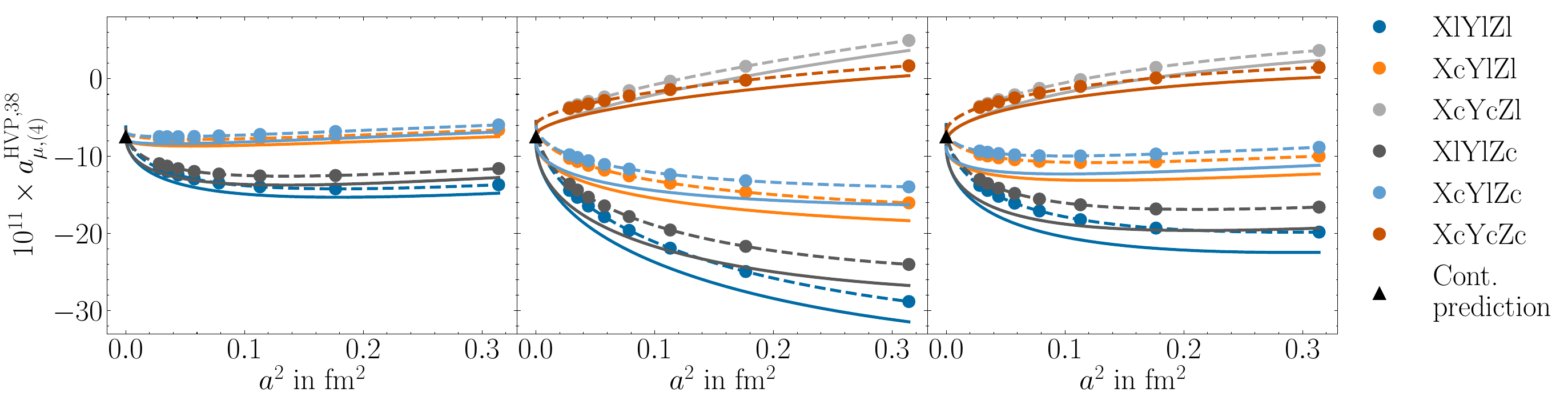}
    \end{subfigure}
    
    \hspace{-1cm}
    \begin{subfigure}{0.19\textwidth}
    \caption{}\label{fig::Gluonless_cont_extra_g}
    \end{subfigure}
    \begin{subfigure}{0.32\textwidth}
    \caption{}\label{fig::Gluonless_cont_extra_h}
    \end{subfigure}
    \begin{subfigure}{0.19\textwidth}
    \caption{}\label{fig::Gluonless_cont_extra_i}
    \end{subfigure}
\caption{Continuum extrapolation of the (4)a diagram, (a)--(c), (4)b diagram (d)--(f) and the sum of them, (4)a + (4)b, (g)--(i), for the `gluon-less' ensembles and $\Lambda=3\, m_\mu$. The three windows are the three different implementations of the photon propagator; left window 2PS method, middle window Fourier method, right window 5D propagator method. The fits are of the form given in eq.~\eqref{equ::Gluonless_Fit_Function}. The dashed lines correspond to fits performed directly on the data points, while the solid lines represent the volume-corrected fits. The different colors correspond to the different discretizations. For (g)--(i) we also include the continuum prediction indicated by a black triangle, which was computed following the methodology outlined in Appendix~A of \cite{Erb:2025nxk}.} 
\label{fig::Gluonless_cont_extra_4b_3m_mu}
\end{figure}

The shape of the integrands of the two diagrams separately is very similar between the 2PS and the 5D propagator method, though more visible differences appear in the sum of the two.
The total integrand of the 5D propagator method is much smoother but also slightly more long-ranged. The total integrand of the Fourier method is also smooth, but has a sign change. This sign change does not appear for some of the other discretizations.

The continuum extrapolation for the (4)a diagram, (4)b diagram and the sum of the two can be seen in figure~\ref{fig::Gluonless_cont_extra_4b_3m_mu}.
We perform a fit to the data points with the function
\begin{align}
    f_{fit}(a)=c_0 + c_1\, a + c_2\, a^2 + c_3 \, a^3. \label{equ::Gluonless_Fit_Function}
\end{align}

The dashed line is the fit directly to the data points, while the solid line includes a volume-correction term. 
We doubled the side lengths of the coarsest lattice and calculated the difference between the two volumes. 
The result is added to the fit to the data points.

One can observe that the extrapolations of the 2PS and 5D propagator method look very similar, with the latter being a bit more spread out. By comparison the extrapolation of the (4)b diagram looks quite different in the Fourier method. 
Figures~\ref{fig::Gluonless_cont_extra_g} to \ref{fig::Gluonless_cont_extra_i} also mark the expected continuum result of $-7.499 \times 10^{-11}$ (see Appendix~B of \cite{Erb:2025nxk}) as a black triangle. 
For all three methods the extrapolations meet up at this point. The results of these extrapolations are collected in table~\ref{tab::Gluonless_cont_extra_Results}. 
For the (4)a and (4)b diagrams separately the results of the extrapolations can be found in tables~\ref{tab::Tables_2PS_gluonless_Results} to \ref{tab::Tables_5D_gluonless_Results} in appendix \ref{app::Tables}.

\begin{figure}[t]
	\centering
	\includegraphics[width=0.49\textwidth]{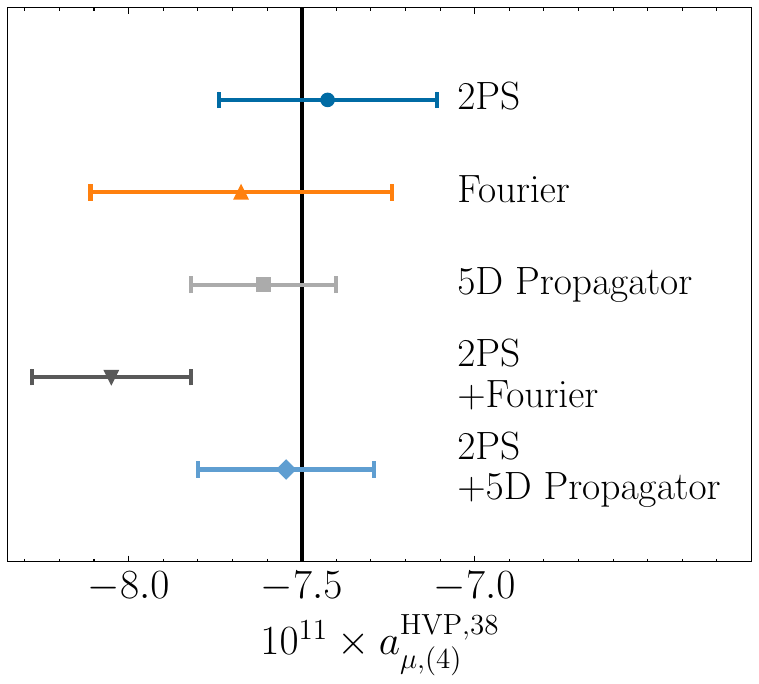}
	\caption{Central value and error from the results in table \ref{tab::Gluonless_cont_extra_Results}. The central values are obtained by taking the mean of the smallest and largest results. The error is half the difference of the same results. The lower two points show a combination of the methods. The second to lowest uses the results of the 2PS method for the (4)a diagram and the results of the Fourier method for the (4)b diagram. The lowest point uses the 5D propagator method instead of the Fourier method. }  
    \label{fig::Gluonless_Error_1}
\end{figure}

We use the spread of the results from the different discretizations to estimate the error of the three different methods. 
For a conservative estimate we use the mean of the smallest and largest results of each method as the central value. 
The error of this central value is then given by half the difference between the smallest and largest results. 
With this setup all results are covered by the error.
The central values with their error are shown in figure~\ref{fig::Gluonless_Error_1} for the three methods. 
They all cover the expected values shown as a black vertical line. 
The spread of the Fourier method is the largest, while the 5D propagator method has the smallest spread.

\begin{table}[b]
    \centering
	\begin{tabular}{c c c c c c c}
		\hline
		 & XlYlZl & XcYlZl &XcYcZl & XlYlZc & XcYlZc & XcYcZc\\ \hline
		2PS & $-7.11$ & $-7.68$ &-& $-7.40$ & $-7.74$ &-\\ \hline
		Fourier & $-7.93$ & $-8.02$ & $-7.24$ & $-8.11$ & $-8.09$ & $-7.27$\\ \hline
		5D propagator & $-7.40$ & $ -7.65$ & $-7.54$ & $-7.67$ & $ -7.82$ & $-7.58$\\ \hline
			
	\end{tabular}
\caption{
    Results of the volume corrected continuum extrapolations in figures~\ref{fig::Gluonless_cont_extra_g} to \ref{fig::Gluonless_cont_extra_i}. The values are given in units of $10^{-11}$. For $\Lambda=3\, m_\mu$ the expected value of the total result is $-7.499 \times 10^{-11}$, as obtained using the methodology outlined in appendix~B of \cite{Erb:2025nxk}.
    }
	\label{tab::Gluonless_cont_extra_Results}
\end{table}

We also show an alternative approach, where we take the results of the (4)a diagram from the 2PS method and the results of the (4)b diagram from either one of the factorization methods. 
The results of these combinations can be found in table~\ref{tab::Tables_Hybrid_Extrapolation_Results} in appendix \ref{app::Tables}.
The combination with the Fourier method is now much further away from the expected value and does not cover it within the error. 
The two discretizations which were on the higher side for the Fourier method are the ones with conserved currents at the $y$-vertex. 
Since these are not available for the 2PS method, only the discretizations with results smaller than the expected value contribute. 
On the other hand, the combination with the 5D propagator method is the closest one to the expected value and has the second smallest errors.

Since this crosscheck was on gluon-less ensembles, where the results on the ensembles are exact, the biggest advantage of the 2PS method, the additional statistics, was not yet relevant, but it will be important in the next section.

\subsection{The connected contribution $\atotvioconn$ on N451 \label{sec::Calculations_QCD_Connected}}
\begin{figure}[t]
		\centering
		\begin{subfigure}{0.49\textwidth}
			\includegraphics[width=\textwidth]{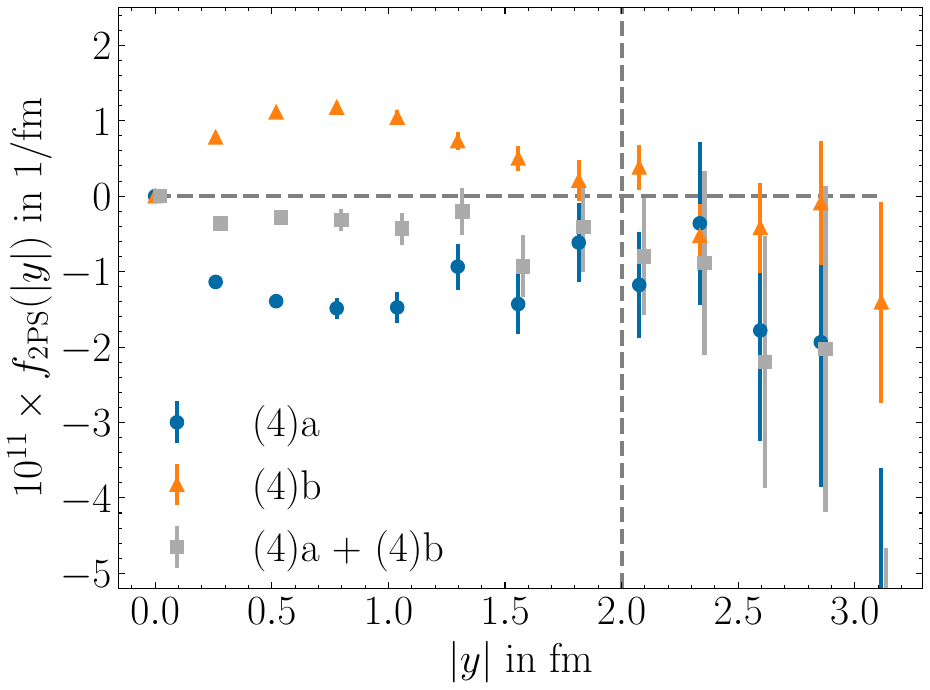}
            \caption{2PS method}
        \label{fig::Conn_QCD_Integrands_2PS}
		\end{subfigure}
		\begin{subfigure}{0.49\textwidth}
			\includegraphics[width=\textwidth]{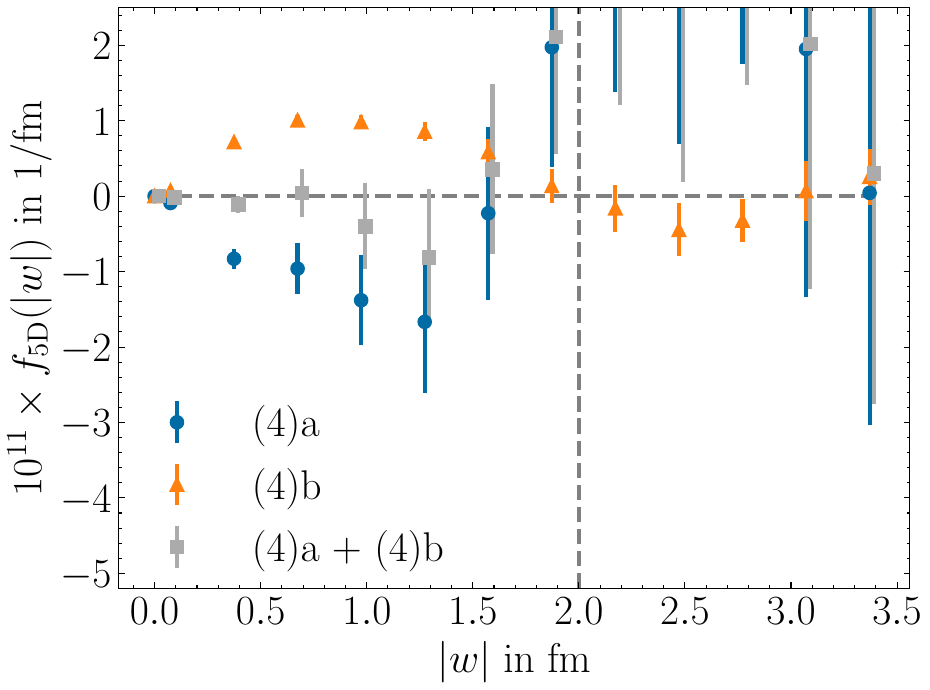}
            \caption{5D propagator method}
        \label{fig::Conn_QCD_Integrands_5D}
		\end{subfigure}
        
		\begin{subfigure}{0.49\textwidth}
			\includegraphics[width=\textwidth]{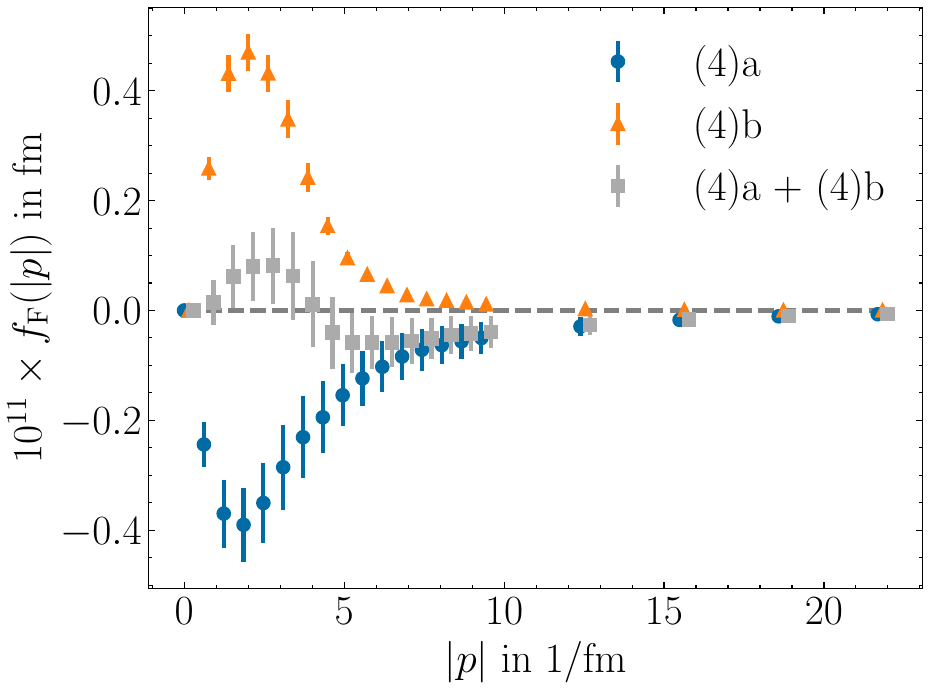}
            \caption{Fourier method}
            \label{fig::Conn_QCD_Integrands_Fourier}
		\end{subfigure}
        \caption{Final integrands of (4)a (blue), (4)b (orange) and the sum of both (gray) on N451 for the three different methods. The PV-mass is $16\,m_\mu$ and the discretization is XcYlZl.}
        \label{fig::Conn_QCD_Integrands}
\end{figure}

With the results of the crosscheck from the previous section, the 5D propagator method seemed to be the most promising. 
It reproduced the continuum result the best.
Now, we want to compare the methods in the context of an ensemble which includes the QCD interactions.
Figure~\ref{fig::Conn_QCD_Integrands} shows the last integrand of the different methods for the fully connected diagrams separately as well as their sum. 
The PV-mass is set to $16\, m_\mu$ and the discretization is XcYlZl. 

All three methods show similar errors for the integrand of the (4)b diagram. 
We can compare the 2PS and 5D propagator method in a bit more detail. 
At small to medium distances the error bars are basically the same between the methods, but at large distances the error bars of the 2PS method drastically increase.
On the other hand the errors of the 5D propagator method show only a small dependence on the distance. 
In the case of the (4)a diagram the advantage of the 2PS method becomes relevant. 
The integrand has much smaller error bars when compared to the other two methods. 
This is because of the factor 48 more statistics of the 2PS method from using the periodicity of the space dimensions, see appendix \ref{app::Diff_4a_2PS_5D}.

\begin{table}[b]
    \centering
    \begin{tabular}{ccccccc}
    \hline
        &XlYlZl & XcYlZl & XcYcZl & XlYlZc & XcYlZc &  XcYcZc\\ \hline
        2PS & - & $-2.12(26)$ & - & - & $-1.95(25)$  & - \\ \hline
        5D & $-1.16(1.40)$ & $-1.24(99)$ & $-1.06(29)$ & $-0.88(1.37)$ & $-1.01(97)$ & $-0.99(29)$ \\ \hline
        Fourier & $-2.41(92)$ & $-1.96(63)$ & $-0.90(13)$ & $-3.33(1.31)$ & $-2.50(90)$ & $-0.66(20)$ \\ \hline
    \end{tabular}
    \caption{Results for the (4)a diagram on N451 for the three different methods. The values are given in units of $10^{-11}$.}
    \label{tab::Conn_QCD_Integral_4a}
\end{table}

\begin{table}[t]
    \centering
    \begin{tabular}{ccccccc}
    \hline
        &XlYlZl & XcYlZl & XcYcZl & XlYlZc & XcYlZc &  XcYcZc\\ \hline
        2PS & - & $1.41(10)$ & - & - & $1.21(10)$  & - \\ \hline
        5D & $1.29(15)$ & $1.27(16)$ & $1.33(17)$ & $1.12(15)$ & $1.10(15)$ & $1.14(16)$ \\ \hline
        Fourier & $1.62(11)$ & $1.66(11)$ & $1.79(12)$ & $1.49(17)$ & $1.51(17)$ & $1.58(17)$ \\ \hline
    \end{tabular}
    \caption{Results for the (4)b diagram on N451 for the three different methods. The values are given in units of $10^{-11}$.}
    \label{tab::Conn_QCD_Integral_4b}
\end{table}

For both the 2PS and 5D propagator method we chose a maximum distance of 2 fm, beyond which we do not evaluate the integrand. 
These larger distances are dominated by noise and do mostly contribute to the error. 
This cutoff is marked by a vertical dashed line in figures~\ref{fig::Conn_QCD_Integrands_2PS} and \ref{fig::Conn_QCD_Integrands_5D}. 
The results of the integrals are written in table~\ref{tab::Conn_QCD_Integral_4a} and \ref{tab::Conn_QCD_Integral_4b}.
In general, the same number of configurations was used across the three methods, except for the Fourier method: 
In the latter case, the discretizations where the current at the $z$-vertex is local use twice as many configurations. 
This is reflected in the errors of the integrals. Where the number of configurations coincides, the errors of the Fourier and 5D propagator method are the same. 
On the other hand with twice more statistics the error is the expected factor of $\sqrt{2}$ smaller. 
For the 2PS method the error of the (4)b diagram is a bit smaller when compared to the other methods, but as expected the error of the (4)a diagram is a lot smaller. 

It is interesting to observe that the variation between the methods is generally larger than the variation between the discretizations of each method. 
The cutoff effects seem to differ more strongly between the methods than between the different discretizations.

\subsection{The disconnected contributions on N451 \label{sec::Calculations_QCD_Disconnected}}
\begin{figure}[t]
		\centering
		\begin{subfigure}{0.49\textwidth}
			\includegraphics[width=\textwidth]{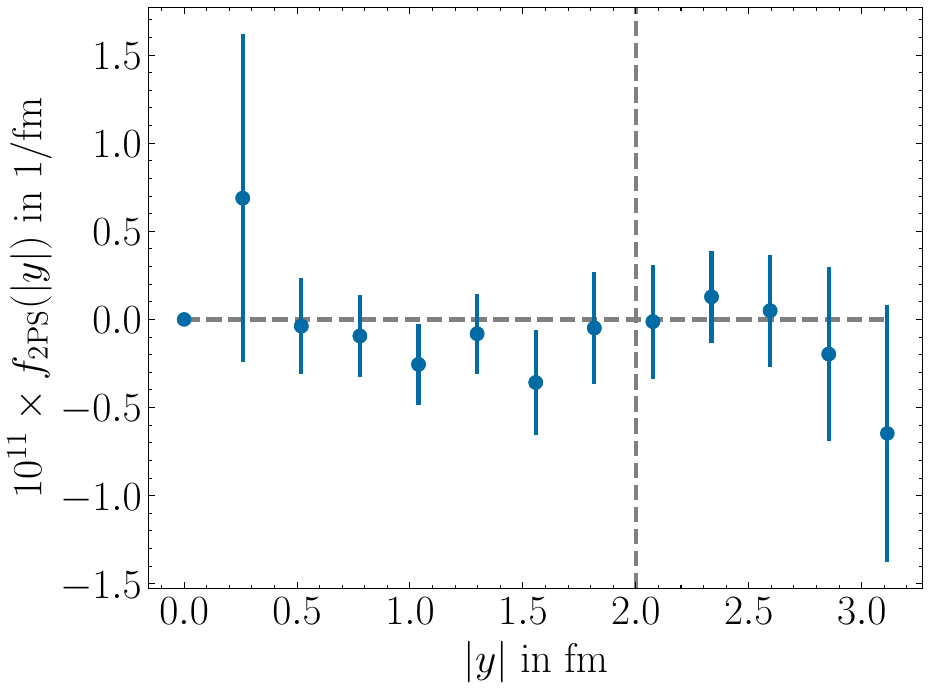}
            \caption{2PS method}
		\end{subfigure}
		\begin{subfigure}{0.49\textwidth}
			\includegraphics[width=\textwidth]{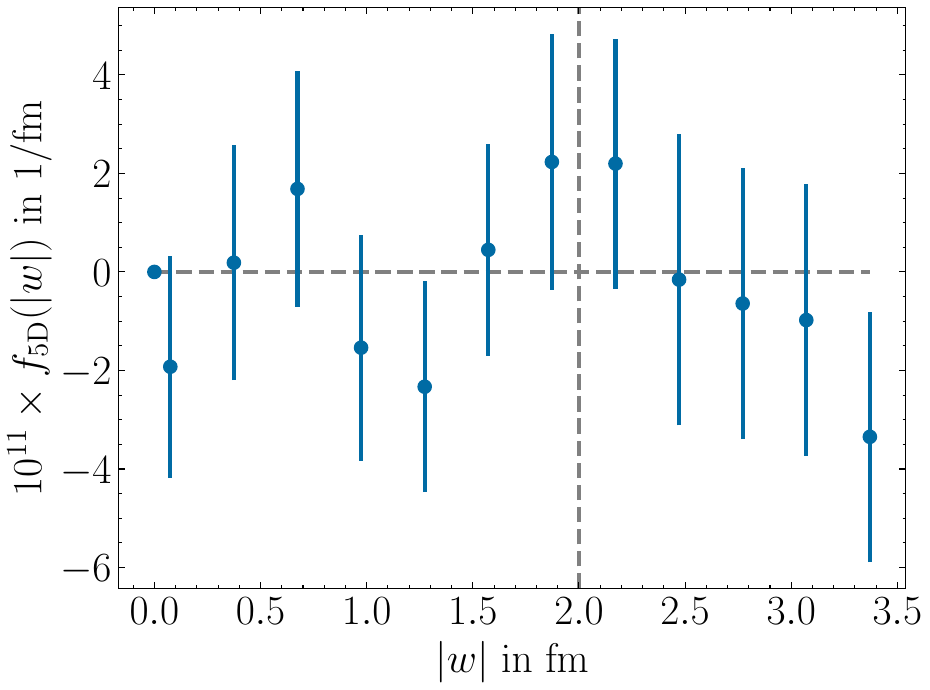}
            \caption{5D propagator method}
		\end{subfigure}
        
		\begin{subfigure}{0.49\textwidth}
			\includegraphics[width=\textwidth]{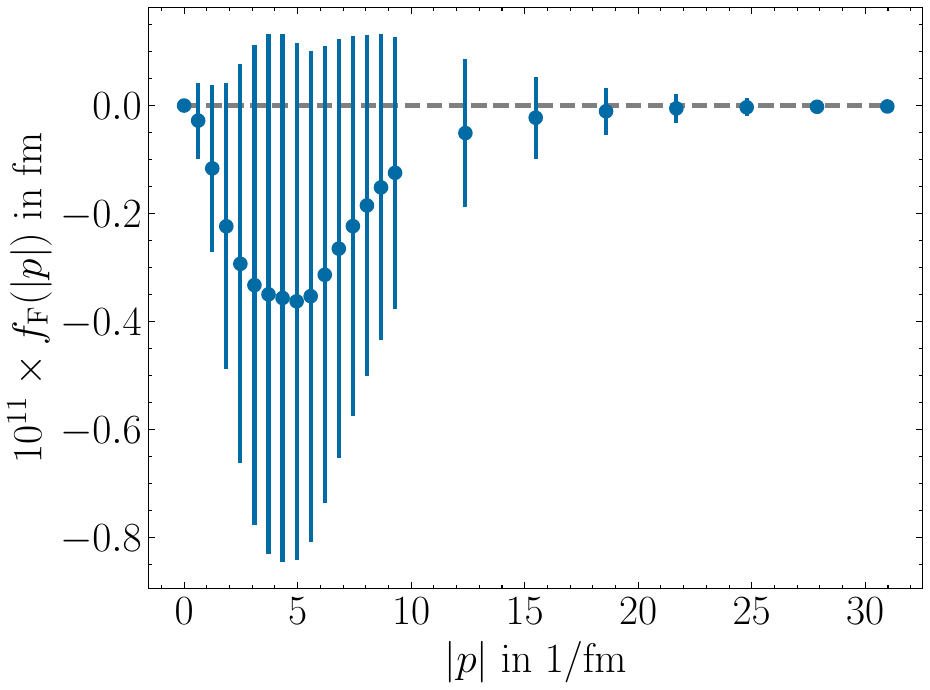}
            \caption{Fourier method}
		\end{subfigure}
        
        \caption{Final integrand of the (3+1)a diagram on N451 for the three different methods. The PV-mass is $16\,m_\mu$ and the discretization is XcYlZl.}
        \label{fig::Disco_3p1a_Integrands}
\end{figure}

\begin{figure}[t]
    \centering
		\begin{subfigure}{0.49\textwidth}
			\includegraphics[width=\textwidth]{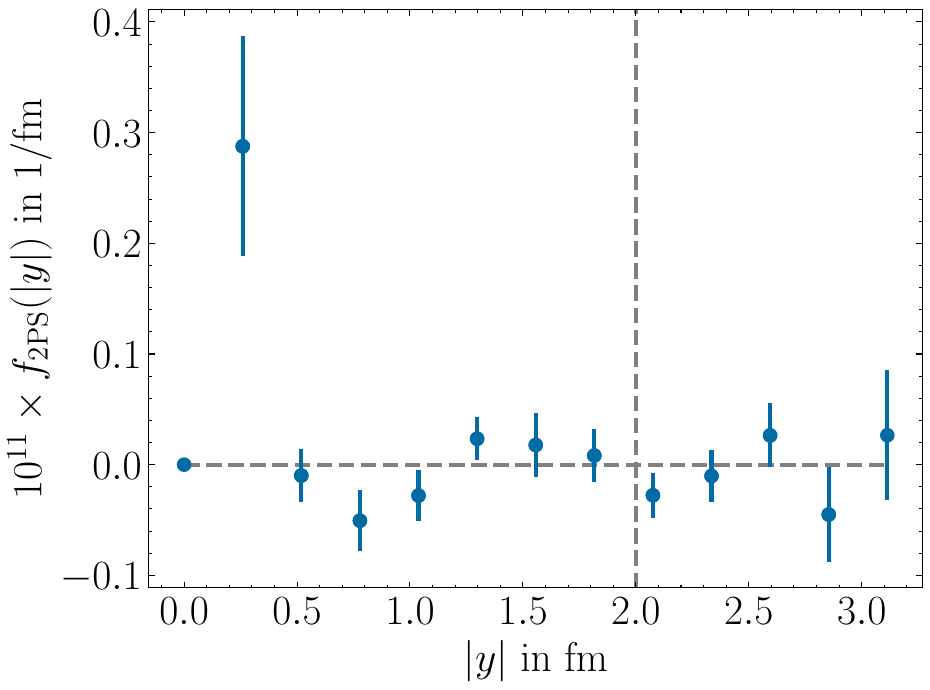}
            \caption{2PS method}
		\end{subfigure}
		\begin{subfigure}{0.49\textwidth}
			\includegraphics[width=\textwidth]{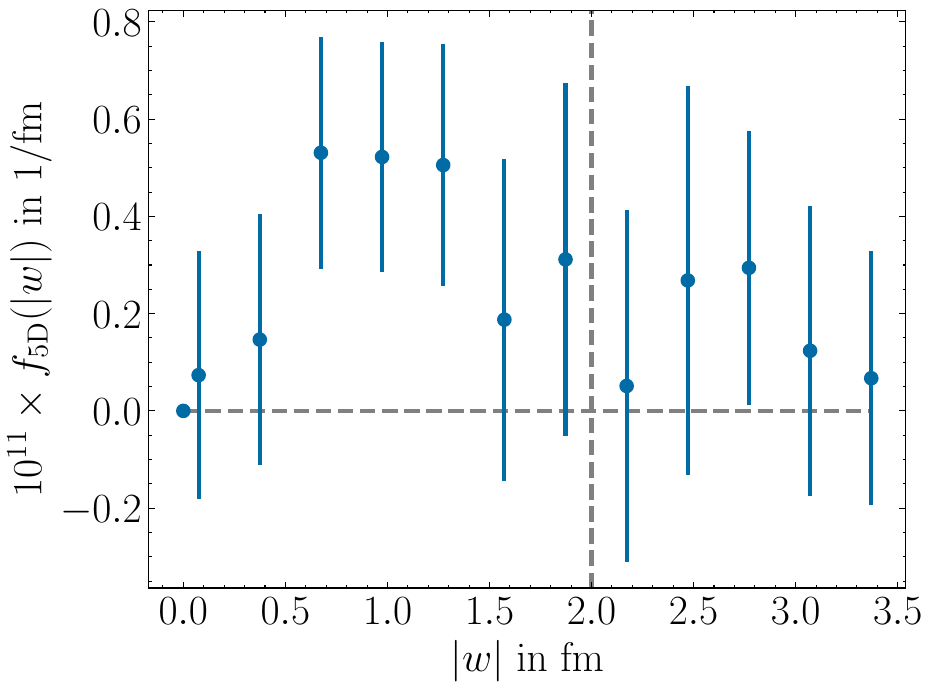}
            \caption{5D propagator method}
		\end{subfigure}
        
		\begin{subfigure}{0.49\textwidth}
			\includegraphics[width=\textwidth]{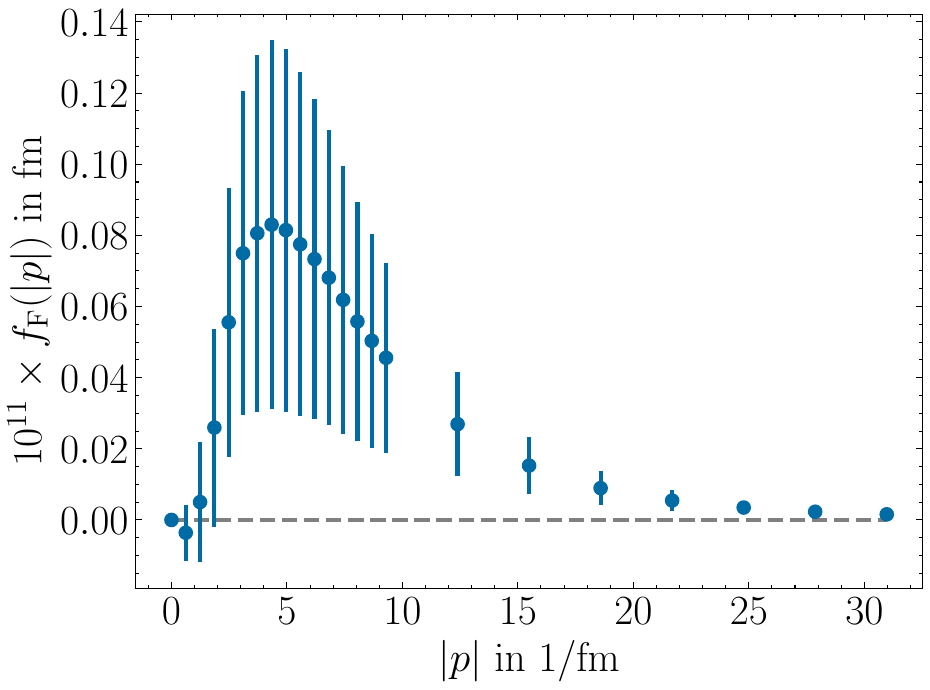}
            \caption{Fourier method}
		\end{subfigure}
        \caption{ Final integrand of the (3+1)a diagram on N451 using the intermediate window version of the CCS kernel for the three different methods. The PV-mass is $16\,m_\mu$ and the discretization is XcYlZl.}
        \label{fig::Disco_3p1a_Window_Integrands}
\end{figure}

In this section we look at the disconnected diagrams, calculated on N451. 
For the first diagram, (3+1)a, we have data for all three methods.
The PV-mass is the same as for the connected diagrams, $\Lambda=16\, m_\mu$.
The final integrands for a discretization of XcYlZl can be seen in figure~\ref{fig::Disco_3p1a_Integrands}. 
The results from the integral are summarized in table~\ref{tab::Disco_Integral_3p1a}. 
For the 2PS and 5D propagator method we imposed a cut of the integral at 2 fm, just as we did for the connected diagrams. 
This cut is mostly motivated by the experience from the connected diagrams, but we also gain additional insight from looking at the intermediate window versions of the integrands. 
The latter can be seen in figure~\ref{fig::Disco_3p1a_Window_Integrands}. 
The 5D propagator method has some points clearly different from zero within error for $|w|$ values that are below the 2~fm cut. 
The results using the intermediate window kernel are provided in table~\ref{tab::Disco_Integral_3p1a_Window}. 

\begin{table}[b]
    \centering
    \begin{tabular}{cccc}
    \hline
        &XlYlZl & XcYlZl & XcYcZl \\ \hline
        2PS & $-0.31(22)$ & $0.09(12) $ & - \\ \hline
        5D & $0.51(4.74)$ & $-0.49(2.96)$ & $-0.79(1.39)$ \\ \hline
        Fourier & $-3.63(8.75)$ & $-2.73(4.45)$ & $-0.16(5.24)$ \\ \hline
    \end{tabular}
    \caption{Results for the (3+1)a diagram on N451 for the three different methods. The values are given in units of $10^{-11}$.}
    \label{tab::Disco_Integral_3p1a}
\end{table}

\begin{table}[t]
    \centering
    \begin{tabular}{cccc}
    \hline
        &XlYlZl & XcYlZl & XcYcZl \\ \hline
        2PS & $-0.032(24)$ & $0.037(13)$ & - \\ \hline
        5D & $0.89(52)$ & $0.63(32)$ & $-0.11(12)$ \\ \hline
        Fourier & $1.59(94)$ & $0.77(47)$ & $0.85(54)$ \\ \hline
    \end{tabular}
    \caption{Results for the (3+1)a diagram on N451 using the intermediate window version of the CCS kernel for the three different methods. The values are given in units of $10^{-11}$.}
    \label{tab::Disco_Integral_3p1a_Window}
\end{table}

The integrand of the Fourier method has some very peculiar behavior. 
The points have large error bars, but they are not scattered in accordance to these errors. 
If one were to ignore the errors, the integrand would look very similar to the integrands of the connected contribution in figure~\ref{fig::Conn_QCD_Integrands_Fourier}.
Indeed, the connected contributions also have this kind of behavior, but it is less pronounced since the errors are much smaller. 
The explanation is given by the large correlation between the points. 
From one point to another there is only a small variation in $p$. 
We could use this observation to calculate fewer momenta, but since each PV-mass needs a slightly different coverage of the integrand we used the number of momenta shown in these plots.

It is clear that for the (3+1)a diagram the 2PS method is the preferred one. 
It has by far the smallest errors, even if one considers the additional statistics from using the periodic diagonal. 
In the full result the error of the 5D propagator method is larger by a factor of 20 and the Fourier method is larger again by almost a factor of two. 
The difference becomes much smaller when looking at the intermediate window version of the kernel. 
But the Fourier method has still much larger errors when compared to the other two methods.

It is also interesting to look at the sign of the integral. 
The full result seems to have a slight tendency to a negative sign, but this comes mostly from the results of the 5D propagator and Fourier methods, where the errors are much larger than the result. 
On the other hand, the intermediate window results show a positive sign, with two exceptions. 
All in all, with our observations here we cannot conclude on a sign or exact value of the (3+1)a diagram, but its contribution seems to be one order of magnitude smaller than the contributions of the connected diagrams. 

\begin{figure}[t]
\centering
		\begin{subfigure}{0.49\textwidth}
			\includegraphics[width=\textwidth]{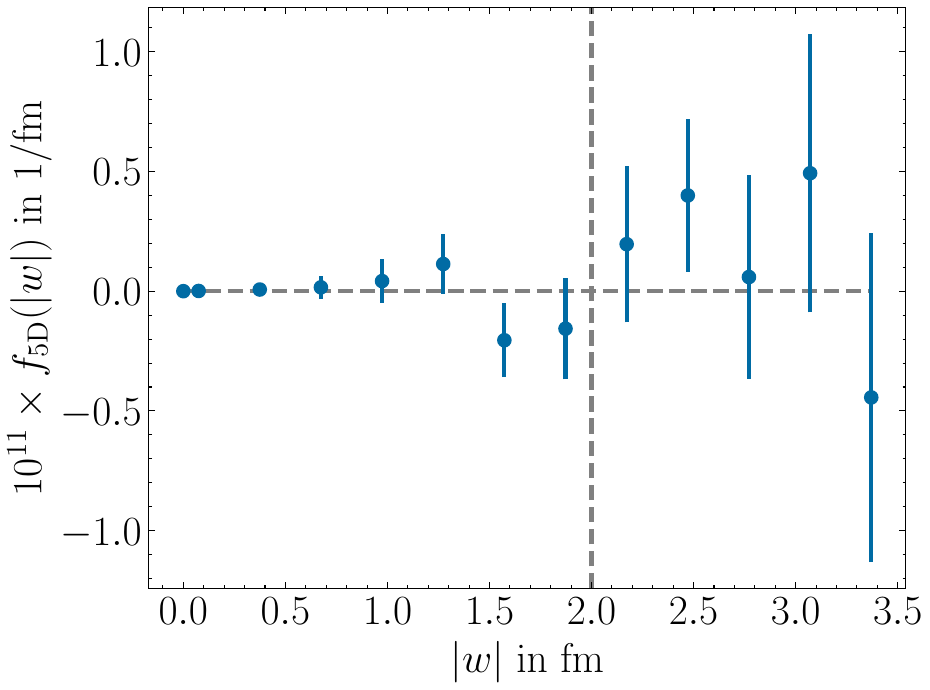}
            \caption{5D propagator method}
		\end{subfigure}
		\begin{subfigure}{0.49\textwidth}
			\includegraphics[width=\textwidth]{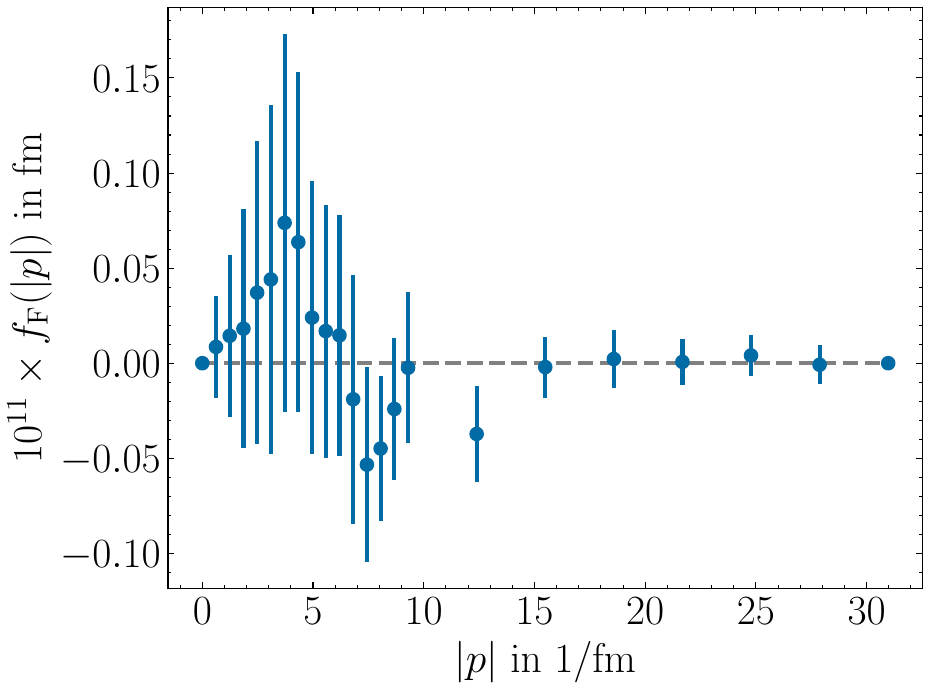}
            \caption{Fourier method}
		\end{subfigure}
        \caption{Final integrand of the (3+1)b diagram on N451 for the two factorization methods. The diagram is UV-finite, so the photon propagator is unregularized. The discretization is XcYlZl.}        
        \label{fig::Disco_3p1b_Integrands}
\end{figure}

For the second diagram, (3+1)b, we have data for the full kernel with the 5D propagator and the Fourier method.
Since it is UV-finite, we present the unregularized results, which amounts to taking the limit $\Lambda \rightarrow \infty$.
The integrands are shown in figure~\ref{fig::Disco_3p1b_Integrands} and the results of the integrals are in table~\ref{tab::Disco_Integral_3p1b}. 
The two factorization methods seem much more suited for this disconnected diagram, as they have much smaller errors when compared to the (3+1)a diagram. 
Yet the Fourier method has still much larger errors than the 5D propagator method. 
For the disconnected diagrams, the suppression of the large distances becomes much more important, because the typical value of the disconnected loop does not by itself fall off with distance. 
The Fourier method does not suppress the large distances, but the 5D propagator method has a suppression of $1/|x|^3$. 
This should explain the difference in the errors between the methods for all the disconnected diagrams. 

The mentioned results of the (3+1)b diagram are zero within error, with a tendency for a negative central value.
Compared to the connected diagrams, the results are at least two orders of magnitude smaller, the same holding true for the errors.

\begin{table}[t]
    \centering
    \begin{tabular}{ccccc}
    \hline
        &XlYlZl & XcYlZl & XlYlZc & XcYlZc \\ \hline
        5D & $-0.06(10)$ & $-0.03(10)$ & $-0.09(10)$ & $-0.06(11)$ \\ \hline
        Fourier & $-0.09(29) $ & $ 0.00(35)$ & $ -0.34(71)$ & $ -3.44(2.57)$ \\ \hline
    \end{tabular}
    \caption{Results for the (3+1)b diagram on N451 for the two factorization methods. The values are given in units of $10^{-11}$.}
    \label{tab::Disco_Integral_3p1b}
\end{table}

\begin{figure}[ht]
\centering
		\begin{subfigure}{0.49\textwidth}
			\includegraphics[width=\textwidth]{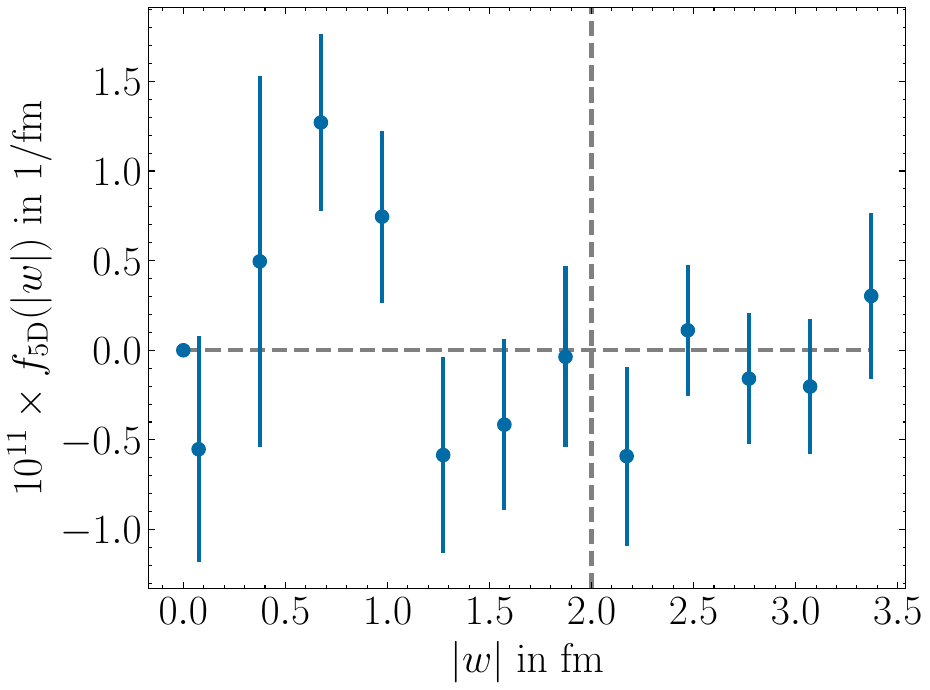}
            \caption{5D propagator method}
		\end{subfigure}
		\begin{subfigure}{0.49\textwidth}
			\includegraphics[width=\textwidth]{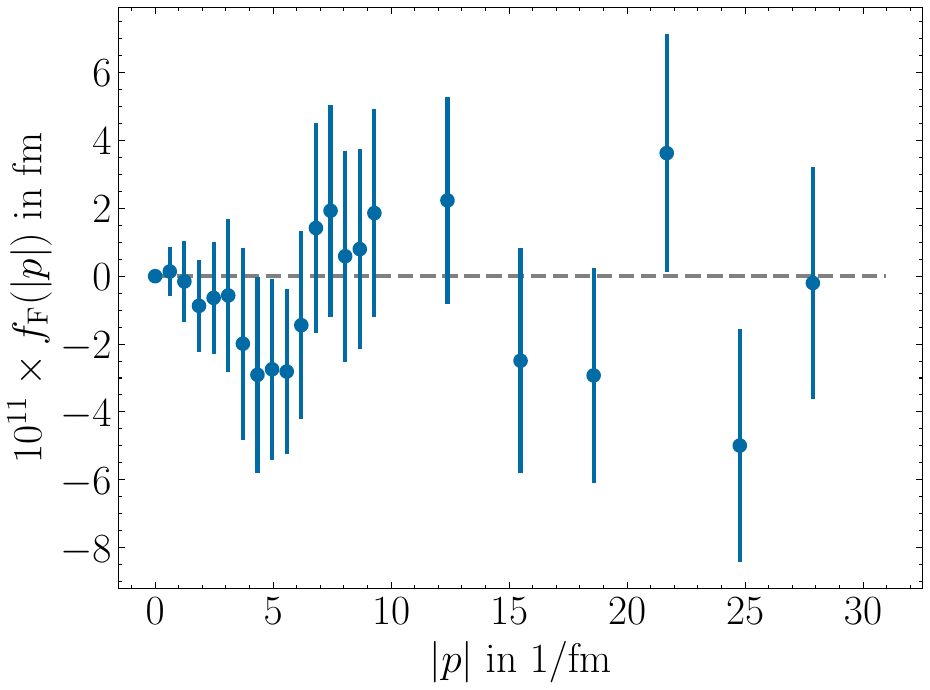}
            \caption{Fourier method}
		\end{subfigure}
        
		\begin{subfigure}{0.49\textwidth}
			\includegraphics[width=\textwidth]{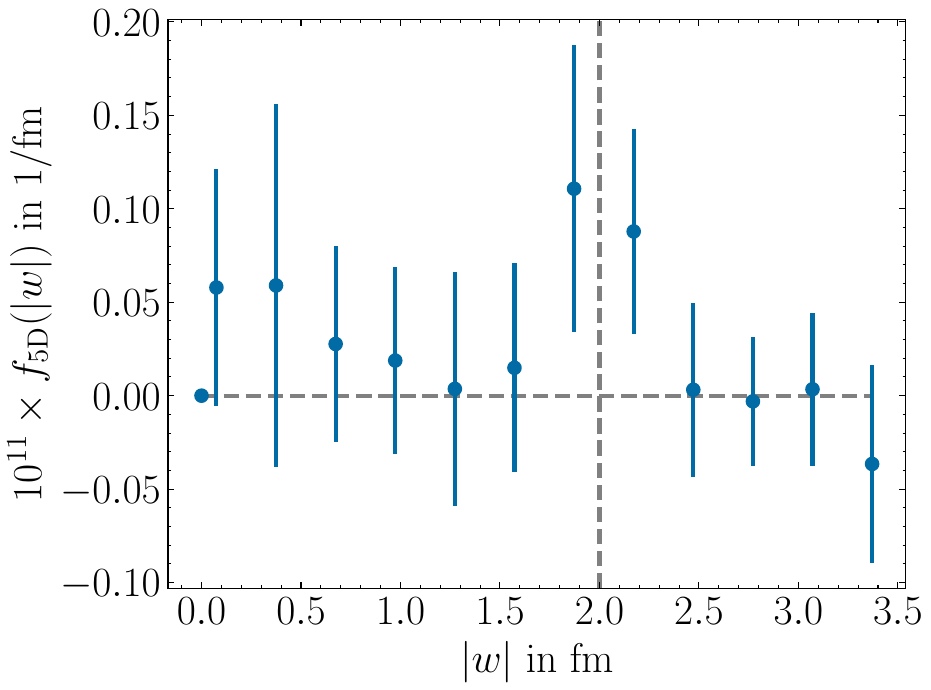}
            \caption{5D propagator method, window kernel}
		\end{subfigure}
		\begin{subfigure}{0.49\textwidth}
			\includegraphics[width=\textwidth]{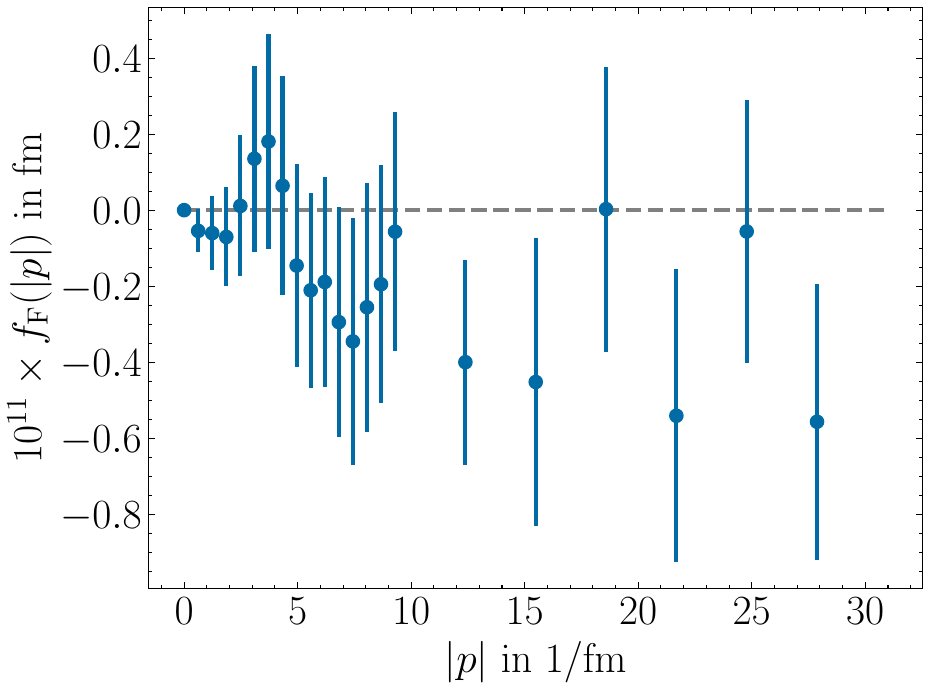}
            \caption{Fourier method, window kernel}
		\end{subfigure}
        \caption{Final integrand of the (2+1+1)b diagram on N451 for the two factorization methods. The upper plots use the full version of the CCS kernel, while the lower plots use the intermediate window version. The diagram is UV-finite, so the photon propagator is unregularized. The discretization is XcYlZl.}        
        \label{fig::Disco_2p1p1b_Integrands}
\end{figure}

The last of the relevant disconnected diagrams is the (2+1+1)b diagram. 
The results we present here were calculated together with the other diagrams.
It should be noted that that a dedicated study would be able to determine the (2+1+1)b diagram to much higher precision, since it only requires the calculation of two-point functions. 
In the future, we will re-use the propagators generated in the 2PS method to achieve this. 
We use the results of this paper to compare the two factorization methods relative to one another. 
The comparison does not include the 2PS method. 
Figure~\ref{fig::Feynman_diagrams_2PS_2p1p1b} shows the approach one would have to take in the calculation using the 2PS method. 
Since we want to sum over both disconnected loops, the propagators in the two point function would be one-to-one propagators, which makes this method unreliable. 

The (2+1+1)b diagram requires an additional step, which was not necessary for the other ones. 
We have to calculate and subtract the vacuum expectation value (VEV). 
The VEV can be obtained by slightly modifying $C^{(2+1+1)b}_{\rho\nu\lambda\sigma}(x, y, z)$ in eq.~\eqref{equ::atotvioD}.
\begin{align}
     C^{(2+1+1)b-VEV}_{\rho\nu\lambda\sigma}(x, y, z)&=-2\, \text{Re} \langle \text{Tr} [ S^l(0, y) \gamma_\nu S^l(y, 0) \gamma_\sigma] \rangle_U \times \label{equ::4pt2p1p1b_VEV}\\
     & \qquad \qquad \qquad \qquad \times \text{Re} \langle\text{Tr} [S^\Delta(x,x)\gamma_\rho ] \times \text{Tr} [S^\Delta(z,z)\gamma_\lambda ] \rangle_U. \nonumber
\end{align}
While the VEV of all other diagrams is equal to zero, for the (2+1+1)b diagram the VEV is much larger and leads to huge cancellations. 

Similar to the (3+1)b we present unregularized results using the 5D propagator method and the Fourier method, but here we also include the intermediate window versions of the CCS kernel. 
The integrands are shown in figure~\ref{fig::Disco_2p1p1b_Integrands}.
For the full CCS kernel the results are gathered in table~\ref{tab::Disco_Integral_2p1p1b} and for the intermediate window version the results are in table~\ref{tab::Disco_Integral_2p1p1b_Window}.

The diagram has a much larger discrepancy between the two methods, even when compared to the other disconnected diagrams.
The results of the 5D propagator method are much more precise than the results form the Fourier method.
Again, this difference comes from the suppression of large distances, which is present in the 5D propagator method but missing in the Fourier method.
Now, there is also a large difference of the error between the two available discretizations. 
We believe this is due to the large VEV of the two-point-quark loop in the case of the XlYlZl discretization, which needs to be subtracted.

\begin{table}[t]
    \centering
    \begin{tabular}{ccc}
    \hline
        &XlYlZl & XcYlZl \\ \hline
        5D & $3.50(2.79)$ & $0.34(58)$ \\ \hline
        Fourier & $9(42)$ & $ -17(29)$ \\ \hline
    \end{tabular}
    \caption{Results for the (2+1+1)b diagram on N451 for the two factorization methods. The values are given in units of $10^{-11}$.}
    \label{tab::Disco_Integral_2p1p1b}
\end{table}

\begin{table}[t]
    \centering
    \begin{tabular}{ccc}
    \hline
        &XlYlZl & XcYlZl \\ \hline
        5D & $0.12(29)$ & $0.06(6)$ \\ \hline
        Fourier & $1.1(4.6)$ & $ -7.2(3.1)$ \\ \hline
    \end{tabular}
    \caption{Results for the (2+1+1)b diagram on N451 using the intermediate window version of the CCS kernel for the two factorization methods. The values are given in units of $10^{-11}$.}
    \label{tab::Disco_Integral_2p1p1b_Window}
\end{table}

\section{Methodological aspects and comparison to the literature \label{sec::Adv_Disadv}}

In this section we first collect the advantages and disadvantages of the different methods.  
We then discuss the computational cost by first calculating the number of inversions of the Dirac operator for each method. 
Thereafter we perform an equal-cost analysis of the errors. 
A comparison of our results with those found in the literature follows. At the end of this section, the idea of using  factorizations more generally is illustrated on the case of the hadronic light-by-light contribution to the muon $(g-2)$.

\subsection{Advantages and disadvantages of the methods}
\subsubsection{2PS method}
The 2PS method has a few important advantages, but they can only be fully realized while calculating the (4)a diagram. Because of this we will first discuss only the calculation of the (4)a diagram and then we will note what changes when also calculating the (4)b diagram. 

The biggest advantage is the use of the periodic diagonal, which gives an additional factor of 48 in statistics in the case of the N451 ensemble. 
We showed that this is the cause of the reduced error of the integral points compared to the results of the other two methods. 
Additionally, since only one-to-all propagators are involved in the calculation, multiple PV-masses can be calculated at the same time. 
It is also possible to access the local and conserved currents at the $x$- and $z$-vertices at no additional cost, but the vertex at the second point source is always a local one. 
Lastly, from a computational point of view this method is very efficient when compared to the factorization methods, since it involves only $12 \times L/2$ inversions of the Dirac operator. 

The disadvantages come into play when also calculating the (4)b diagram. It involves a sequential propagator, which now includes the photon propagator. 
This drastically reduces the gain from the periodicity of the diagonal, as the sequential propagator has to be calculated for each position of the $y$-vertex. 
Since the photon propagator is included, calculating multiple PV-masses now also involves additional inversions of the Dirac operator. 
Similarly, if one wants to have both a conserved and a local current at the $x$-vertex, twice as many sequential propagators have to be calculated. 

There is also a significant disadvantage, unrelated to the (4)b diagram, that we have not yet addressed properly.
To use the periodicity we are storing all of the one-to-all propagators on the diagonal. 
Therefore we need a lot of memory for our calculation. 
This scales super-linearly with the spatial extent of the ensembles, because not only does the needed memory for each propagator get larger, but the number of stored propagators also grows.
However, there are ways to get around this limitation, for example by further increasing the step size or by not storing all of the propagators and recalculating some. 

One last minor disadvantage comes from the fact that we do not automatically have access to the unregularized version of the photon propagator. 
For the connected diagrams this is not an issue, but some of the disconnected diagrams are UV-finite. 
This last disadvantage can be completely removed by also including a contraction with the unregularized version of the photon propagator during the calculation.

\subsubsection{Fourier method}

While the Fourier method has some big advantages, it also comes with significant disadvantages. 
From the three different methods it has the smallest initial cost, 
although this is not as big of a difference as one might expect it to be, because the integrand changes more rapidly, when compared to the integrands of the other two methods. 
It is also the case that you need slightly different coverage for different PV-masses.
These two facts together increases the number of momenta one needs to cover. 
The big advantage comes into play when doing the last integral. 
It is possible to adjust the photon propagator in this step.
That means one can test any PV-mass, has access to the unregularized version of the photon propagator and one could even test other regularization schemes. 

The disadvantage arises from the weight functions for the internal vertices. 
Since they are given by an exponential function with a purely imaginary argument, they do not fall off at large distances. The noise from the points at these distances is not suppressed. 
In the calculations we presented, this has little to no effect for the connected diagrams. 
But for the disconnected diagrams, the Fourier method consistently showed much larger errors when compared to the other two methods.
It  may also explain why the Fourier method was significantly further away from the expected result in the gluon-less calculation than the other two methods.

\subsubsection{5D propagator method}

Of the three methods, the 5D propagator method has the largest initial cost. 
Since the summands of the PV-regulated photon propagator get split separately an additional factor of three appears. 
The methods is also worse on N451 for the (4)a diagram, when compared to the 2PS method, as it lacks the option of using the periodicity of the diagonal. 

For the (4)b diagram the advantages of the 5D propagator method give it an edge. 
The errors also only slightly depend on the distance to the origin. 
This is because the weight function of the internal vertices falls off with at least $1/x^3$, suppressing the noise from large distances. 
The suppression also benefits the results of the disconnected diagrams, where the 5D propagator method shows consistently smaller errors than the Fourier method.
Splitting the calculation of the photon propagator also has two advantages. 
It gives access to the unregularized version, which is of interest for the UV-finite subset of the disconnected diagrams. 
Additionally, the cost of further PV-masses is reduced, as the unregularized part can be reused from the previous calculations.

As a last point, one should keep in mind that the 5D propagator method reproduced the expected result in the gluon-less calculation the best, adding confidence in its applicability.

\subsection{Computational cost}
In the following, we list the number of inversions of the Dirac operator which are necessary for each method. 
In all three cases one connected diagram depends on the inversions which were done for the other connected diagram. 
In the 2PS method the calculation of the (4)b diagram relies on the inversions from the calculation of the (4)a diagram, while for the Fourier and 5D propagator method it is the other way around. 
One can view one diagram a generating the base cost of the method and the other diagram being the add-on. 
We get the disconnected diagrams without any additional inversions from the base cost in all methods. 

For each calculation of a propagator we inverted the different spin and color combinations separately.
Thus, for one propagator, twelve inversion of the Dirac operator are necessary.
Another factor that concerns all methods is the possibility of different combinations of local and conserved currents. 
It usually costs more to include additional discretizations.
If there is a factor which could be reduced by calculating fewer combinations, it will be denoted with a $d$ (for discretizations) in the subscript.

For readability we use the abbreviation $L$ for the spatial extent of the ensemble, $N_{PV}$ for the number of PV-masses and $N_M$ for the number of momenta in the Fourier method. 
For the 2PS method there is also the abbreviation $N_{TS}$ for the number of points on the diagonal used as an origin when calculating the (4)b diagram. 
These are the sequential propagators for which we use truncated solves. They are faster by about a factor of three when compared to the untruncated solves. 

In the next subsections, we will discuss the concrete numbers for each method. 
They are summarized in table~\ref{tab::Dirac_inv}, which also includes how many inversions are necessary for one PV-mass on the N451 ensemble, where $L=48$  and $N_M=22$. 

Lastly, we calculated an equal cost comparison of the errors between the methods. 
It can be found in table~\ref{tab::Cost_scaled_Errors}. 
For a fair comparison, we assumed that three PV-masses are calculated with each method. 

\subsubsection{2PS method}
For calculating the (4)a diagram the 2PS method needs only the point sources on the diagonal, so $12 \times L/2$ inversions of the Dirac operator. 
Multiple PV-masses and discretizations can be calculated at the same time with no additional cost. 
On the N451 ensemble 288 inversions are needed for this diagram.

For the (4)b diagram the number of inversions per chosen origin increases by multiple factors. 
Most importantly the sequential propagators have to be calculated for each position of the $y$-vertex, giving a factor of $(L/2-1)$ for the number of inversions.
Additionally, we have a factor of $N_{PV}$ from the inclusion of the photon propagator in the sequential propagator, a factor of $4$ from the inclusion of the gamma matrix and a factor of $2$, if one is interested in discretizations where the current at the $x$-vertex is local as well as conserved.
The inversions are done for $N_{TS}$ points on the diagonal with less precise solves and one time with exact solves to calculate the bias correction. 

On the N451 ensemble this gives in total for one PV-mass $1104 \times 2_d$ exact inversions. 
The number of inversions with truncated solves is $4416\times 2_d$ for $N_{TS}=4$. 
The speedup of the truncated solves compared to the exact one is about a factor of three. 
So the number of truncated inversions is equivalent to $1472 \times 2_d$ exact inversions.

\subsubsection{Fourier method}
In the Fourier method the calculation of the (4)b diagram forms the baseline. 
It needs one inversion for the first one-to-all propagator. 
The number of inversions for the sequential propagators have the factor $N_M$ for the momentum which has to be included and a factor of 4 from the gamma matrix. 
There is also an additional factor of 2 from the different discretizations. 
On N451 the base line is given by $1068 \times 2_d$ number of inversions.

The added-on inversions for the (4)a diagram have the same factors, but instead of the factor of 2 for the discretizations it is a factor of 3. 
So analogous to before the additional inversions on N451 are $1056 \times 3_d$.

One last thing to keep in mind for the Fourier method is the independence from the PV-mass. 
This means no diagram includes a factor of $N_{PV}$.

\begin{table}[t]
\centering
    \begin{tabular}{ c c r}
         Diagram (4)a & & \\ \hline
        2PS & $\boldsymbol{6\times L}$ & $\boldsymbol{288}$\\ \hline
        Fourier & $ 48\times N_{ M} \times 3_{ d}$ & $1056\times 3_d$ \\ \hline
        5D Propagator & $ 36 \times L  \times N_{PV} \times 3_d$ & $1728^* \times 3_d$ \\ \hline
    &&\\
         Diagram (4)b & &\\ \hline
        2PS & Ex.: $48 \times (L/2 -1) \times N_{ PV} \times 2_{ d}$ & $1104^* \times 2_d$\\ 
            &  Tr.: $48/3 \times N_{TS} \times(L/2 -1) \times N_{ PV} \times 2_{ d}$& $1472^* \times 2_d$ \\ \hline
        Fourier & $\boldsymbol{12\times (1+ N_M \times 4 \times 2_d})$ & $\boldsymbol{1068 \times 2_d}$ \\ \hline
       5D Propagator & $\boldsymbol{12\times (1+ L \times 3 \times N_{PV} \times 2_d)}$ & $\boldsymbol{1740^* \times 2_d}$ \\ \hline
    \end{tabular}

    \caption{Number of necessary inversions of the Dirac operator for each method.  The subscript $d$ marks factors which can be reduced to 1, if less discretizations are calculated. Bold font marks the numbers of inversions which are at least necessary, i.e. the calculations with normal font rely on the results of the ones with bold font, but not the other way around. In each row after the general form there is also the number of inversions necessary for the N451 ensemble for one PV-mass. If this number increases for multiple PV-masses it is marked by a star. The factor of 1/3 for the truncated inversions accounts for the gained speedup.}
    \label{tab::Dirac_inv}

\end{table}

\subsubsection{5D propagator method}
The 5D propagator method is very similar to the Fourier method but with two key differences. 
Instead of a factor of $N_M$ we have a factor of $L/4$, because we go up to half the spatial extent of the ensemble and only consider every second point.
The inversions now also depend on the PV-mass, which has two effects. 
First of all we have the factor of 3, since the three terms of the regularized photon propagator have to be calculated separately. 
Secondly, both diagrams have a factor of $N_{PV}$, because the photon propagator is included in the inversions for the sequential and double sequential propagators. 

In total the base line number of inversions for the 5D propagator method on N451 for one PV-mass is $1740\times 2_d$ and the additional number of inversions for the (4)a diagram is $1728\times 3_d$.
For subsequent PV-masses this gets reduced to at least $1164 \times 2_d$ and $1152 \times 3_d$ respectively, because at least the zero mass term only has to be computed once.

\subsubsection{Equal-cost comparison of the errors}

Determining the number of inversions of the Dirac operator for each method is useful, but for a meaningful comparison we also have to take into account the resulting errors. 
Comparing the calculations of all diagrams between the methods is straightforward. 
Using the core-hours each method needed to compute one configuration, we can calculate the time-equivalent number of configurations between the methods. 
As a reference point we used the 250 configurations calculated with the Fourier method for all discretizations where the $z$-vertex is local.
The resulting scaled errors can be seen in rows 2, 4 and 6 of table~\ref{tab::Cost_scaled_Errors}.
For all results in the table, we assumed three PV-masses and the maximum number of discretizations are calculated.

There are also two further groups for comparisons noted in the table. 
For the first one we compare the errors with equal cost, if only one of the fully connected diagrams is calculated. 
To compute the core hours one would need for one configuration in this setup we simply subtracted the time needed for the additional (double) sequential propagators. 
This is only an approximation of the real time it would take, but for our purposes here this comparison is sufficient. 
The scaled errors for this group comparisons are in rows 3, 5 and~7.

The physics application requires that we compute both fully connected diagrams. 
The last two lines account for that by having the scaled errors, if the 2PS method is used for the (4)a diagram and one of the factorization methods is used for the (4)b diagram.

\begin{table}[t]
    \centering
    \begin{tabular}{c c c c c c c}
    \hline
         & \#Configurations & (4)a & (4)b & (3+1)a & (3+1)b & (2+1+1)b\\ \hline
         Fourier Full& 250 & $0.63$ & $0.11$ & $4.45$& $0.35$ & $29$\\ \hline
         Fourier (4)b only& 373 & - & $0.091$ & $3.65$& $0.29$ & $24$\\ \hline
         2PS Full& 51& $0.41$ & $0.16$ & $ 0.19$ & - &-\\ \hline
         2PS (4)a only& 420& $0.15$ & - & $ 0.066$ & - &-\\ \hline
         5D Propagator Full& 72 & $1.31$ & $0.22$ & $3.91$& $0.14$ & $0.77$\\ \hline
         5D Propagator (4)b only& 105 & - & $0.18$ & $3.23$& $0.11$ & $0.64$\\ \hline
         2PS (4)a + 5D (4)b& 87 & $0.32$ & $0.20$ & $0.15$& $0.12$ &$0.70$\\ \hline
         2PS (4)a + Fourier (4)b& 198 & $0.21$ & $0.13$ & $0.096$& $0.40$ &$33$\\ \hline
    \end{tabular}
    \caption{Equal-cost comparison between the methods and further combinations under the assumption of calculating three PV-masses and the maximum number of discretizations. We show the errors for the XcYlZl discretization from all diagrams, scaled to the equal cost number of configurations. 
    }
    \label{tab::Cost_scaled_Errors}
\end{table}

\subsection{Comparison to the literature}

The current literature on isospin-breaking effects is still sparse, which makes comparisons difficult, especially when looking at the disconnected diagrams. 
The BMW collaboration provides a partial breakdown of their results in their 2020 publication \cite{Borsanyi:2020mff}. 
These results are renormalized in the BMW scheme, defined by
\begin{align}
    m_\pi = 134.9768(5) \text{ MeV}, \quad M_{\rm ss} = 689.89(49) \text{ MeV}, \quad w_0 = 0.17236(70) \text{ fm}, 
\end{align}
where $M_{\rm ss}$ is the meson with two mass-degenerate quarks with the mass of the strange quark and $w_0$ is computed from the gradient-flowed gauge field \cite{BMW:2012hcm}.
Their results are also already in the continuum and at physical pion mass. 
Therefore one has to be careful with the comparisons. 

Table~\ref{tab::Comparison_BMW} shows their results together with the appropriate combinations of diagrams from this work with adjusted charge factors.
The factors are different, because they used electromagnetic currents at the external vertices. 
The errors from this work for the first two combinations are smaller when compared to their results, but this may change when the couterterms are included, the photon regulator is removed and after the continuum extrapolation. 
The combination in the second-to-last line with only the (3+1)b diagram leads to the same result for both BMW and this work, but the error of this work is larger. 
Recall that the diagram is UV-finite, so that no counterterms and photon regulators are necessary to evaluate it. 
The results for the (2+1+1)b diagram, which is also UV-finite, have central values of opposite signs; but in both cases the error is much larger than the value. 

\begin{table}[t]
    \centering
    \begin{tabular}{cccc}
    \hline
        Diagrams & Charge factors em & BMW-20  & This work \\ 
        &&($m_\pi =135$ MeV)&($m_\pi =280$ MeV)\\ \hline
        (4)a + (4)b & 17/81 & $–12.3(4.0)_{\rm stat}(3.1)_{\rm syst}$ & $-6.42(2.35)$\\ \hline
        (2+2)a + (3+1)a & 25/81 + 7/81 & $–5.5(1.5)_{\rm stat} (1.0)_{\rm syst}$ &  $ -6.28(1.27)$ \\ \hline
        (3+1)b & 7/81 & $–0.093(86)_{\rm stat}(95)_{\rm syst}$ & $-0.093(343)$\\ \hline
        (2+1+1)b & 5/81 & $–0.11(24)_{\rm stat}(14)_{\rm syst}$ & $0.38(65)$\\ \hline
    \end{tabular}
    \caption{Comparison between the values from \cite{Borsanyi:2020mff} and the results of this work scaled with the appropriate charge factor. Values are given in units of $10^{-11}$.}
    \label{tab::Comparison_BMW}
\end{table}

\subsection{More general factorizations: the case of $a_\mu^{\rm HLbL}$}

The aspect of factorizing sums over vertices in lattice QCD correlation functions is not restricted to such vertices directly connected by a photon propagator. 

We take as an example the hadronic light-by-light scattering in the muon $(g-2)$. Starting from Eqs.\ (6.1--6.3) in the WP'25~\cite{Aliberti:2025beg}, and making use of translation invariance and current conservation, we arrive at
\begin{align}
    a_\mu^{\rm HLbL} = &
    -\frac{e^6}{48 m_\mu} \int d^4x \,d^4y\,d^4z\,z_\rho \langle j_\sigma(z)\,j_\mu(x) \,j_\nu(0)\,j_\lambda(y)\rangle
 \\ &   \int d^4u\,d^4v\,d^4w\, G_0(w-x) G_0(u) G_0(y-v) \,e^{-ip\cdot (w-v)}
\nonumber \\ & \qquad    {\rm Tr}\Big\{ 
[\gamma_\rho,\gamma_\sigma]
    (-ip\!\!\!/ + m_\mu) \gamma_\mu S(w-u) \gamma_\nu S(u-v) \gamma_\lambda (-ip\!\!\!/ + m_\mu)
    \Big\},
    \nonumber
\end{align}
where $p=im_\mu\hat p$, $\hat p\in\mathbb{R}^4$, $\hat p^2 = 1$, $S(x)= 2\pi x\!\!\!/  \,G_{m_\mu}^{(2)}(x)+m_\mu G_{m_\mu}^{(1)}(x)$ is the muon propagator (see Eq.\ (\ref{eq:Gmlda}) for the explicit form of $G_m^{(\lambda)}(x)$), and $G_0(x)= 1/(4\pi^2 x^2)$ the massless scalar propagator. Defining the function
\begin{align}
    L^{f}(p,x) &=
    \int d^4w\; G_0(x-w)\, e^{-ip\cdot w}\, S(w),
\end{align}
which can be expressed as a one-dimensional integral over a Feynman parameter, we can express $a_\mu^{\rm HLbL}$ in the following way,
\begin{align}
    & a_\mu^{\rm HLbL} = 
    -\frac{e^6}{48 m_\mu} \int d^4u\, G_0(u)\,  \int d^4x \,d^4y\,d^4z\,z_\rho \langle j_\sigma(z)\,j_\mu(x) \,j_\nu(0)\,j_\lambda(y)\rangle
 \\ &  \quad\times  {\rm Tr}\Big\{ 
[\gamma_\rho,\gamma_\sigma]
    (-ip\!\!\!/ + m_\mu) \gamma_\mu L^{f}(p,x-u)\gamma_\nu  L^{f}(p,u-y) \gamma_\lambda (-ip\!\!\!/ + m_\mu)
    \Big\}.
    \nonumber
\end{align}
It shows that, for fixed $u$, the weight functions for the vertices $x$ and $y$ factorize (up to the Dirac trace, which can be reduced to a trace over $2\times 2$ matrices). For the connected diagrams contributing to the QCD four-point function of the electromagnetic current, this observation allows one to perform the sums ($\sum_{x,y,z}$) over three out of four vertices using the technique of sequential propagators. Whether such an approach is superior to the one that was pursued in the existing calculations~\cite{Chao:2021tvp,Blum:2023vlm,Fodor:2024jyn,Kalntis:2025imd}, where lattice-wide sums are performed only over two of the four vertices but only point sources were needed, remains to be seen. We note that averaging over the unit vector $\hat p$, a technique applied in~\cite{Asmussen:2022oql}, is not possible without entangling the integrals over $x$ and $y$. Thus the integral over $u$, which is to be carried out last, can  be reduced to a two-dimensional integral, but not to a one-dimensional one.
\section{Conclusion \label{sec::Conclusion}}

In the context of computing electromagnetic corrections to hadronic vacuum polarization in the muon $(g-2)$ using lattice QCD, we have compared three different implementations of the infinite-volume photon propagator with a Pauli-Villars (PV) style regularization. 
The implementations are the two-point-source (2PS) method, and two methods where we factorized the dependence on the relative coordinates of the two internal vertices connected by a photon propagator.
The factorization methods considered are the Fourier method and the 5D propagator method.
For the comparisons, we used the calculation of the isospin-violating part of the hadronic vacuum polarization (HVP), which corresponds to the cross-terms in the two-point correlation function of the electromagnetic current $j_\mu^{em}$, when the latter is written as the sum of an isovector ($j_\mu^3$) and an isoscalar ($j_\mu^8$) current.
We computed the two contributing connected diagrams as well as three of the four disconnected diagrams.

The PV regularization enables us to do crosschecks between lattice calculations on `gluon-less' ensembles and continuum calculations. 
We obtained results consistent with the continuum calculation with all three implementations of the photon propagator. 
For further comparisons, we computed the relevant diagrams on a CLS gauge ensemble (N451) with a pion mass of 286~MeV. 
In all cases the results of the three methods are consistent with each other, but there are differences in the errors. 
Considering only the connected diagrams and comparing only the errors, at equal cost the 2PS method is the preferred one for the (4)a diagram and the Fourier method is the preferred one for the (4)b diagram.

But there are two important points, which promote the 5D propagator method to be the preferred one for the (4)b diagram. 
The first one comes from the gluon-less crosscheck, where the Fourier method had the largest spread of all the methods. 
Additionally, when combined with the 2PS method in the hybrid approach, the mean value with its error did not cover the predicted value. 
This is in contrast to the 5D propagator method, which had the smallest spread on its own and was the closest one to the predicted value when combined with the 2PS method.
The other point comes from the disconnected diagrams. 
The Fourier method gives larger errors for them compared to the 5D propagator method, likely because it does not suppress the large distances.
In contrast, the 5D propagator method suppresses the large distances with at least $1/|x|^3$, which gives it an advantage over the Fourier method in this case.
We expect the difference in efficiency between the methods to increase as the pion mass is lowered.
Note that the 5D-propagator factorization formula also applies to a lattice-regularized photon propagator, up to a very short-range correction --- see appendix~\ref{sec:latprop}.

Based on these observations, we will use a hybrid method going forward.
The (4)a diagram will be computed by the 2PS method and the (4)b diagram will be computed by the 5D propagator method. 
The disconnected diagrams can be calculated at no additional cost with both methods, making it possible to choose the method afterward, based on the errors. 
Adopting this approach, we obtain on ensemble N451 a total result of
\begin{align}
    2\,\atotvioemN (\Lambda=16\, m_\mu) &=(-\overset{(4)a}{4.24(52)}+\overset{(4)b}{2.54(32)} - \overset{(2+2)a-ll}{1.18(22)} + \overset{(3+1)a}{0.09(12)} - \overset{(3+1)b}{0.06(20)}) \times 10^{-11} \nonumber\\
    &=  -2.85(70) \times 10^{-11}. \label{equ::atotvioemN_result}
\end{align}
Note that all diagrams except (3+1)a include an additional factor of two to account for the equally sized (3,8) and (8,3) contributions. 
We also included the light-light contribution of the (2+2)a diagram, which was calculated in \cite{Parrino:2025afq}. 
The full result of the (2+2)a diagram would also include the light-strange contribution, but it is more than an order of magnitude smaller than the light-light contribution and vanishes within the error.
The two UV-finite diagrams, (2+2)a and (3+1)b, use the unregularized version of the photon propagator, which amounts to taking the limit $\Lambda \rightarrow \infty$.
For the other diagrams, the Pauli-Villars mass is set to $\Lambda=16\, m_\mu$.
For the results shown here, we used the discretization with a conserved current at the $x$-vertex and local currents everywhere else, since this combination is available for all methods and diagrams. Our calculation, though not performed at physical quark masses, confirms the extremely small size of the (3+1)-topology diagrams relative to the connected diagrams, as found in Ref.~\cite{Borsanyi:2020mff}.

The (2+1+1)b diagram is omitted from the total result for now. 
It is zero within error, but the error is of the same size as the one of the (4)a diagram. 
Its precision in this paper was limited by the statistics of the other diagrams. 
A dedicated study would be able to determine it with much smaller errors. 
In fact, re-using the propagators generated by the 2PS method would already give it a factor $L/2$ more statistics, where $L$ is the spatial extend of the ensemble. 
We will use this optimization in our calculations going forward.
It should be noted that the charge factor of the (2+1+1)b diagram relative to the connected diagrams is enhanced by more than a factor of six in the (3,8) contribution as compared to the full HVP contribution.

The result in eq.~\eqref{equ::atotvioemN_result} is larger by almost a factor of two compared to the appropriate intermediate result of our previous publication at the SU(3)$_{\rm f}$ symmetric point.
This confirms the growing importance of this contribution when going towards the physical point.
Still, its size is much smaller than the error of the current experimental determination of $a_\mu$, which is $14.5\times 10^{-11}$, and the expected value of the strong isospin-breaking contribution at the physical point, which amounts to $30.0(7.5)\times 10^{-11}$ according to the model estimate of~\cite{Erb:2025nxk}.

\acknowledgments 
H.M.\ thanks Xu~Feng for discussions on the evaluation of the connected diagrams.
We acknowledge the support of Deutsche Forschungsgemeinschaft (DFG, German Research Foundation) through project HI~2048/1-3 (project No.~399400745), project JRP (No.~458854507) of the research unit FOR 5327 “Photon-photon interactions in the Standard Model and beyond exploiting the discovery potential from MESA to the LHC”, and through the Cluster of Excellence “Precision Physics, Fundamental Interactions and Structure of Matter” (PRISMA+ EXC 2118/1) funded within the German Excellence Strategy (project ID 39083149). 
One of the authors (K.O.) has been supported by DFG through project HI~2048/1-3 (project No.~399400745).
Calculations for this project were partly performed on the HPC cluster and “HIMster II” at the Helmholtz-Institut Mainz and “Mogon II” at JGU Mainz. 
The authors gratefully acknowledge the computing time made available to them on the high-performance computer "Mogon NHR Süd-West" at the NHR Center JGU Mainz. This center is jointly supported by the Federal
Ministry of Research, Technology and Space and the state governments participating in the National High-Performance Computing (NHR) joint funding program (http://www.nhr-verein.de/en/our-partners). We are grateful to our colleagues in the CLS initiative for sharing ensembles.
All plots were created using the matplotlib library \cite{Hunter:2007}.\\

\appendix

\section{Charge factors \label{app::Charge_factors}}
We calculate the relevant charge factors for each diagram of figure \ref{fig::Formalism_feynman} in this appendix.
The charge matrices are defined in eq.~\eqref{equ::charge_matrices}.
We will also use the notation $\text{Tr}_l$ and $\text{Tr}_s$ to note if a trace only goes over the light sector or the strange quark, respectively. 
The modified traces can be related to the usual trace via
\begin{align}
\text{Tr}_l\{\mathcal{Q}\}=\text{Tr}\{\mathcal{Q}^{(l)} \mathcal{Q} \},& \quad
\mathcal{Q}^{(l)} = \begin{pmatrix}
    1 & 0&0 \\
    0& 1 &0 \\
    0 & 0 & 0
\end{pmatrix},\\
\text{Tr}_s\{\mathcal{Q}\}=\text{Tr}\{\mathcal{Q}^{(s)} \mathcal{Q} \},& \quad
\mathcal{Q}^{(s)} = \begin{pmatrix}
    0 & 0&0 \\
    0& 0 &0 \\
    0 & 0 & 1
\end{pmatrix}.
\end{align}
We first look at the fully connected diagrams, (4)a and (4)b. 
They have no contribution from the strange quark. 
For both diagrams the charge factor is simply given by
\begin{align}
        \fQ^{(4)a}=\fQ^{(4)b}&= \text{Tr} \{ \mathcal{Q}^{(3)}\mathcal{Q}^{(em)}\mathcal{Q}^{(em)}\mathcal{Q}^{(8)} \}= \frac{1}{36}. \label{equ::CF_chargefactors_4a}
\end{align}
The (2+2)a diagram has two possible flavor combinations, which result in two possible charge factors.
In the first one both loops consist only of light quarks, while in the second one the loop with the isovector current has a strange quark and the isoscalar current has a light quark.
Their charge factors are
\begin{align}
    \fQ^{(2+2)a-ll}&=\text{Tr}_l \{ \mathcal{Q}^{(3)}\mathcal{Q}^{(em)}\} \text{Tr}_l \{\mathcal{Q}^{(em)}\mathcal{Q}^{(8)} \} = \frac{1}{36},\\    
    \fQ^{(2+2)a-ls}&=\text{Tr}_l \{ \mathcal{Q}^{(3)}\mathcal{Q}^{(em)}\} \text{Tr}_s \{\mathcal{Q}^{(em)}\mathcal{Q}^{(8)} \} = \frac{1}{18}.
\end{align}
At the SU(3)$_{\rm f}$ symmetric point, where $m_l=m_s$, both charge factors have to be added together giving a total charge factor of 1/12.

The (3+1)a and (3+1)b diagrams each have one one-point loop, where the quark can either be a light quark or a strange quark. 
So each of them has two charge factors,
\begin{align}
    \fQ^{(3+1)a-ll}&=\text{Tr}_l\{ \mathcal{Q}^{(3)}\mathcal{Q}^{(em)}\mathcal{Q}^{(em)} \} \text{Tr}_l \{ \mathcal{Q}^{(8)} \}=\frac{1}{18},\\
    \fQ^{(3+1)a-ls}&=\text{Tr}_l\{ \mathcal{Q}^{(3)}\mathcal{Q}^{(em)}\mathcal{Q}^{(em)} \} \text{Tr}_s \{ \mathcal{Q}^{(8)} \}=-\frac{1}{18},\\
    \fQ^{(3+1)b-ll}&=\text{Tr}_l\{ \mathcal{Q}^{(3)}\mathcal{Q}^{(em)}\mathcal{Q}^{(8)} \} \text{Tr}_l \{ \mathcal{Q}^{(em)} \}=\frac{1}{36},\\    
    \fQ^{(3+1)b-ls}&=\text{Tr}_l\{ \mathcal{Q}^{(3)}\mathcal{Q}^{(em)}\mathcal{Q}^{(8)} \} \text{Tr}_s \{ \mathcal{Q}^{(em)} \}=-\frac{1}{36}.
\end{align}
The charge factors of the different flavor combinations only differ in the sign. 
If follows directly that at the SU(3)$_{\rm f}$ symmetric point the charge factors are equal to 0.

In eq.~\eqref{equ::formalism_SDelta} we introduce the notation $S^\Delta(x,y)$, which is the difference between the light quark propagator and the strange quark propagator, $S^\Delta(x,y)=S^l(x,y)-S^s(x,y)$. 
It already encodes the sign difference we see here. 
So for our notation the flavor combinations of the (3+1) diagrams are combined and each diagram has a single charge factor,
\begin{align}
    \fQ^{(3+1)a}&=\frac{1}{18},\\
    \fQ^{(3+1)b}&=\frac{1}{36}.
\end{align}
Lastly, the (2+1+1)b diagram has four flavor combinations, but the same observations as with the (3+1) diagrams hold true. 
The four charge factors are
\begin{align}
    \fQ^{(2+1+1)b-lll}&=\text{Tr}_l \{ \mathcal{Q}^{(3)}\mathcal{Q}^{(em)}\} \text{Tr}_l \{ \mathcal{Q}^{(em)} \}\text{Tr}_l \{ \mathcal{Q}^{(8)} \}=\frac{1}{18},\\
    \fQ^{(2+1+1)b-lls}&=\text{Tr}_l \{ \mathcal{Q}^{(3)}\mathcal{Q}^{(em)}\} \text{Tr}_l \{ \mathcal{Q}^{(em)} \}\text{Tr}_s \{ \mathcal{Q}^{(8)} \}=-\frac{1}{18},\\
    \fQ^{(2+1+1)b-lsl}&=\text{Tr}_l \{ \mathcal{Q}^{(3)}\mathcal{Q}^{(em)}\} \text{Tr}_s \{ \mathcal{Q}^{(em)} \}\text{Tr}_l \{ \mathcal{Q}^{(8)} \}=-\frac{1}{18},\\
    \fQ^{(2+1+1)b-lss}&=\text{Tr}_l \{ \mathcal{Q}^{(3)}\mathcal{Q}^{(em)}\} \text{Tr}_s \{ \mathcal{Q}^{(em)} \}\text{Tr}_s \{ \mathcal{Q}^{(8)} \}=\frac{1}{18}.    
\end{align}
At the SU(3)$_{\rm f}$ symmetric point they add up to 0 and using the $S^\Delta(x,y)$ notation we can simplify them to a single charge factor
\begin{align}
    \fQ^{(2+1+1)b}&=\frac{1}{18}.
\end{align}

\section{Derivation of the factorization formula for the scalar propagator \label{app::Derivation_5D}}

In this appendix, we provide a derivation of the representation of the continuum scalar propagator used in section \ref{sec:5dpropmeth}.
We also provide the derivation of an analogous result for the scalar propagator on an infinitely extended lattice.

\subsection{The continuum case}

In this appendix, we derive the representation ($x,y\in\mathbb{R}^d$, $d=2\lambda+2$)
\begin{align}
  \la{eq:GmFacto}
G_{m}^{(\lambda)}(x-y) = 4 \int d^dw\;  G_m^{(\lambda+1/2)}(0,w-y)\,G_m^{(\lambda+1/2)}(0,x-w)
\end{align}
of the scalar propagator in $d$-dimensional Euclidean space, in which the $x$ and the $y$ dependence is factorized. By making the change of integration variables $w=v+ y$, the representation is equivalent to having a convolution of two $(d+1)$-dimensional propagators,
\begin{align}
  \la{eq:GmConvol}
G_{m}^{(\lambda)}(x-y) = 4 \int d^dv\;  G_m^{(\lambda+1/2)}(0,v)\,G_m^{(\lambda+1/2)}(0,x-y-v).
\end{align}

We proceed via momentum space and derive a more general result (eq.\ (\ref{eq:UneqMass}) below) for the convolution of two unequal-mass propagators. Labelling the Cartesian coordinate of the `extra dimension' with index zero, we may start from the mixed representation 
\begin{align}
 G_m^{(\lambda+1/2)}(b,v) &= \int_{-\infty}^{\infty} \frac{dp_0}{2\pi} e^{ip_0 b}\int \frac{d^dp}{(2\pi)^d}  \frac{e^{ip\cdot v}}{p_0^2 +  p^2+m^2} 
\nonumber\\ &=  \int_{-\infty}^\infty \frac{dp_0}{2\pi} e^{ip_0 b}\, G_{\sqrt{p_0^2+m^2}}^{(\lambda)}(v) \qquad (b\in\mathbb{R},v\in\mathbb{R}^d)
\end{align}
of the $(d+1)$-dimensional propagator.
Making use of this representation for both propagators, one easily finds the following result for the Fourier transform,
\ba
4\int d^dz \int d^dv\, G_M^{(\lambda+1/2)}(b,v)\,G_m^{(\lambda+1/2)}(b,z-v)\,e^{-ikz}
= \frac{e^{-|b|(\sqrt{k^2+m^2}+\sqrt{k^2+M^2})}}{\sqrt{(k^2+m^2)(k^2+M^2)}}\,.
\ea
Setting $b=0$ and inverting the Fourier transform, we have
\ba
4 \int d^dv\, G_M^{(\lambda+1/2)}(0,v)\,G_m^{(\lambda+1/2)}(0,z-v)
&=& \int \frac{d^dk}{(2\pi)^d} \frac{e^{ikz}}{\sqrt{(k^2+m^2)(k^2+M^2)}}\;.
\ea
Now noting that
\be
\frac{1}{\sqrt{(k^2+m^2)(k^2+M^2)}} = \frac{1}{\pi} \int_{-\pi/2}^{\pi/2} d\theta\;\frac{1}{k^2 +\frac{1}{2}(M^2+m^2) + \frac{1}{2}(M^2-m^2)\sin\theta}\;,
\ee
one finds
\be\label{eq:UneqMass}
4 \int d^dv\, G_M^{(\lambda+1/2)}(0,v)\,G_m^{(\lambda+1/2)}(0,z-v) = \frac{1}{\pi} \int_{-\pi/2}^{\pi/2} d\theta\;G^{(\lambda)}_{\sqrt{\frac{1}{2}(M^2+m^2) + \frac{1}{2}(M^2-m^2)\sin\theta}}(z).
\ee
In the equal mass case $M=m$, the right-hand side immediately simplifies to $G^{(\lambda)}_m(z)$, which proves eq.\ (\ref{eq:GmConvol}) upon setting $z=x-y$.

\subsection{The case of lattice propagators\label{sec:latprop}}

We consider propagators on the infinite $d$-dimensional cubic lattice $\Lambda_d$ with lattice spacing $a$.
Let $\hat p_\mu \equiv \frac{2}{a}\sin \frac{ap_\mu}{2}$ and $\hat p^2 \equiv \sum_{\mu=1}^d (\hat p_\mu)^2$ for $p\in\mathbb{R}^d$.
We recall the expression of the standard one-dimensional lattice propagator,
\begin{align}
    \la{eq:Gm_minushalf}
\hat G_m^{(-1/2)}(a,x_1=a\cdot n_1) 
&\equiv \int_{-\pi/a}^{\pi/a} \frac{dp_1}{2\pi}\,\frac{e^{ip_1x_1}}{\hat p_1^2+m^2}
\nonumber\\
&= \frac{1}{m\sqrt{4+ a^2m^2}} \left(\frac{2}{2+a^2m^2+am\sqrt{4+ a^2m^2}}\right)^{|n_1|}.
\end{align}
Note that the $(d+1)$-dimensional propagator is related to the $d$-dimensional one in an analogous way to the continuum case,
   \begin{align}
\hat G_m^{(\lambda+1/2)}(b,v) &= \int_{-\pi/a}^{\pi/a} \frac{dp_0}{2\pi} e^{ip_0 b}\int_{-\pi/a}^{\pi/a} \frac{d^dp}{(2\pi)^d}  \frac{e^{ip\cdot v}}{\hat p_0^2 + \hat p^2+m^2} 
\nonumber\\ &=  \int_{-\pi/a}^{\pi/a} \frac{dp_0}{2\pi} e^{ip_0 b}\,\hat G_{\sqrt{\hat p_0^2+m^2}}^{(\lambda)}(v).
\end{align}
 Labeling the `extra-dimension' coordinate as component zero of the position-space vector, we compute the convolution of two $(d+1)$-dimensional propagators on the infinite $d$-dimensional lattice, 
\[
   a^d \sum_{v\in\Lambda_d} \hat G_M^{(\lambda+1/2)}(b,v)\,\hat G_m^{(\lambda+1/2)}(b,z-v).
   \]
   Similarly to the continuum case, we Fourier-transform the expression,
   \ba
&& a^d \sum_z  a^d \sum_{v} \hat G_M^{(\lambda+1/2)}(b,v)\,\hat G_m^{(\lambda+1/2)}(b,z-v) \,e^{-ik\cdot z}
 \nonumber  \\ &=& \int_{-\pi/a}^{\pi/a} \frac{dp_0}{2\pi}\,e^{ip_0b}\,a^d \sum_v \hat G_M^{(\lambda)+1/2)}(b,v)\,\frac{e^{-ik\cdot v}}{\hat k^2 + \hat p_0^2+m^2}
  \nonumber \\ &=& \hat G^{(-1/2)}_{\sqrt{\hat k^2+m^2}}(b) \;a^d \sum_v e^{-ik\cdot v}  \int_{-\pi/a}^{\pi/a} \frac{dq_0}{2\pi} \,e^{iq_0b}\,\hat G^{(\lambda)}_{\sqrt{\hat q_0^2+M^2}}(v)
   \\ &=& \hat G^{(-1/2)}_{\sqrt{\hat k^2+m^2}}(b)  \int_{-\pi/a}^{\pi/a} \frac{dq_0}{2\pi} \,e^{iq_0b}\, \frac{1}{\hat k^2 + \hat q_0^2+M^2}
\nonumber   \\ &=& \hat G^{(-1/2)}_{\sqrt{\hat k^2+m^2}}(b) \;\hat G^{(-1/2)}_{\sqrt{\hat k^2+M^2}}(b).
 \nonumber
 \ea
 The Fourier transform is thus given by the product of two massive one-dimensional lattice propagators, in complete analogy with the continuum case.
 
   We continue by setting $b=0$ and $M=m$; in that case (see eq.\ (\ref{eq:Gm_minushalf})),
   \ba
   && a^d \sum_z  a^d \sum_{v} \hat G_m^{(\lambda+1/2)}(b,v)\,\hat G_m^{(\lambda+1/2)}(b,z-v) \,e^{-ik\cdot z}
  \nonumber \\ &=& \frac{1}{\hat k^2+m^2} \; \frac{1}{4+a^2\hat k^2 + a^2m^2}
    = \frac{1}{4}\left[\frac{1}{\hat k^2+m^2} - \frac{a^2}{4+a^2\hat k^2 + a^2m^2} \right].
   \ea
   Thus we recognize the expression for the difference of two momentum-space propagators and conclude
   \be
  4\, a^d \sum_{v\in\Lambda_d} \hat G_m^{(\lambda+1/2)}(0,v)\,\hat G_m^{(\lambda+1/2)}(0,z-v)
   = \hat G_m^{(\lambda)}(z) - \hat G_{\sqrt{m^2+4/a^2}}^{(\lambda)}(z).
   \ee
   Setting $z=x-y$ and making the change of summation variable $v=w-y$,
   we can factorize a lattice propagator, up to a very short-range term,
   \begin{align}
   & \hat G_m^{(\lambda)}(x-y) 
   =  \hat G_{\sqrt{m^2+4/a^2}}^{(\lambda)}(x-y)
\\ & \qquad +4\, a^d \sum_{w\in\Lambda_d} \hat G_m^{(\lambda+1/2)}(0,w-y)\,\hat G_m^{(\lambda+1/2)}(0,x-w).
  \nonumber   \end{align}
     The first term on the right-hand side vanishes exponentially beyond a distance of a few lattice spacings. Therefore its contribution to a correlation function can presumably be evaluated with low statistics.
\section{2PS method without using periodicity \label{app::Diff_4a_2PS_5D}}

\begin{figure}[ht]
\centering
		\begin{subfigure}{0.49\textwidth}
			\includegraphics[width=\textwidth]{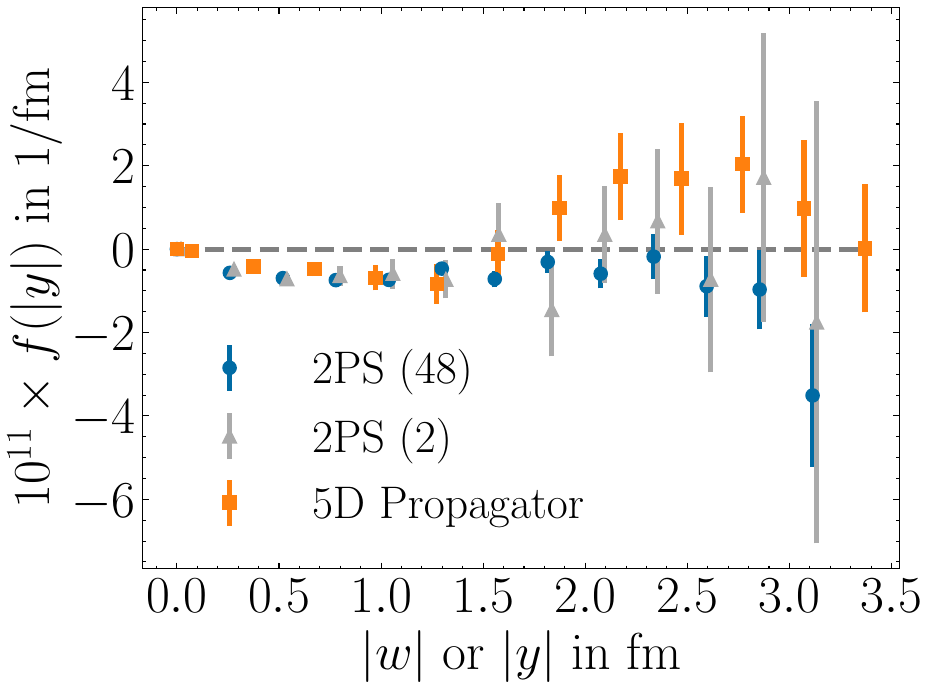}
            \caption{Full scale}
            \label{fig::2PSvs5D_4a_Full_scale}            
		\end{subfigure}
		\begin{subfigure}{0.49\textwidth}
			\includegraphics[width=\textwidth]{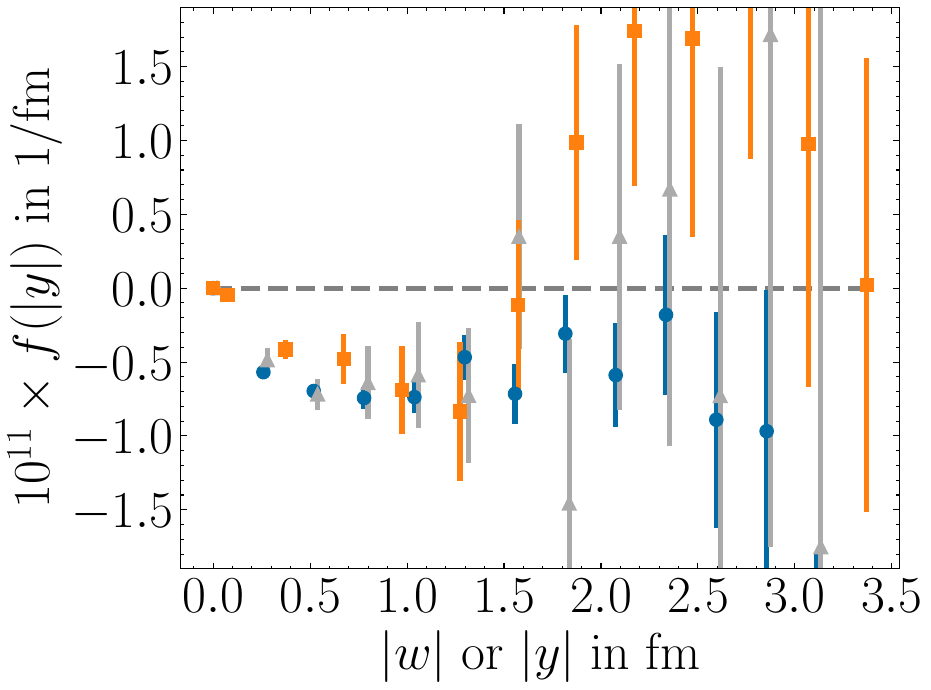}
            \caption{Zoomed in}
            \label{fig::2PSvs5D_4a_Zoomed_in}
		\end{subfigure}
        \caption{Comparison of the final integrand of diagram (4)a between the 2PS method (blue dots), the 2PS method with less statistics (gray triangles) and the 5D propagator method (orange squares). The gray triangles are shown with a slight offset to make them better distinguishable from the blue dots.}        
        \label{fig::2PSvs5D_4a}
\end{figure}    

During the calculation of the (4)a diagram with the 2PS method we were able to use the periodicity in space of the ensemble to gain a factor of 48 in statistics. 
In section \ref{sec::Calculations_QCD_Connected} we claimed this additional statistics is the reason for the small errors of the 2PS method when compared to the other two methods.
Here we address this claim. 

Figure~\ref{fig::2PSvs5D_4a} depicts the final integrand of the calculation of the (4)a diagram for both the 2PS method and the 5D propagator method. 
The plot is shown two times. 
Once on the left side, with vertical axis at full scale to make comparisons at large distances easier, and once on the right side, with the vertical axis zoomed in to make comparisons at small distances easier. 
It is important to keep in mind that the horizontal axis is different between the two methods, which means comparing them this way is not completely rigorous, but the observations from this appendix hold true regardless. 

The blue dots show the 2PS method at full statistics, using the additional factor of 48. 
They are the same as the blue dots in figure~\ref{fig::Conn_QCD_Integrands_2PS}. 
The gray triangles are the results of the 2PS method, but with only one of the points on the diagonal used as a source. 
Because of the forward and backward symmetry there is still a factor of two more statistics compared to the 5D propagator method, which is noted in the legend of figure~\ref{fig::2PSvs5D_4a_Full_scale}.
Lastly, the orange squares are the results of the 5D propagator method. They are the same as the blue dots in figure~\ref{fig::Conn_QCD_Integrands_5D}.

We already observed in the main body of the text, that the 2PS method at full statistics has generally smaller error bars at the relevant distances below 2 fm.
Here, we also see that the errors of the 5D propagator method have only a slight dependence on the distance, just like one can see it in figure~\ref{fig::Conn_QCD_Integrands_2PS} for the (4)b diagram. 
But the main observation is given by the comparison between the 2PS method with reduced statistics and the 5D propagator method. 
Below 1.5 fm the errors of them are around the same size but beyond that point the 5D propagator method as an advantage. 
From this we conclude that the much smaller errors of the 2PS method in table~\ref{tab::Conn_QCD_Integral_4a} truly is from the gain in statistics from using the periodicity.

\section{Tables \label{app::Tables}}
Tables (\ref{tab::Tables_2PS_gluonless_Results},~\ref{tab::Tables_Fourier_gluonless_Results},~\ref{tab::Tables_Fourier_gluonless_Results},~\ref{tab::Tables_Hybrid_Extrapolation_Results}) in this appendix summarize our results obtained from  `gluonless' lattices (i.e.\ all gauge link variables $U(\mu,x)$ are set to unity).

\begin{table}[ht]
	\centering
	\begin{tabular}{c c c c c c c}
		\hline
		 & XlYlZl & XcYlZl &XcYcZl & XlYlZc & XcYlZc & XcYcZc\\ \hline
		Total & $-7.11$ & $-7.68$ &-& $-7.40$ & $-7.74$ &-\\ \hline
		2-Loop & $\ 28.66$ & $\ 28.36$ &-& $\ 28.56$ & $\ 28.32$ &-\\ \hline
		Self-Energy & $-35.77$ & $-36.04$ &-& $-35.96$ & $-36.07$ &-\\ \hline
			
	\end{tabular}
	\caption{Results of continuum extrapolation on gluonless ensembles for two-point-source method with $\Lambda =3\, m_\mu$. The values are given in units of $10^{-11}$.
    }
	\label{tab::Tables_2PS_gluonless_Results}
\end{table}

\begin{table}[ht]
\centering
	\begin{tabular}{c c c c c c c}
		\hline
		 & XlYlZl & XcYlZl &XcYcZl & XlYlZc & XcYlZc & XcYcZc\\ \hline
		Total & $-7.93$ & $-8.02$ & $-7.24$ & $-8.11$ & $-8.09$ & $-7.27$\\ \hline
		2-Loop & $\ 27.95$ & $\ 27.86$ &$\ 27.90$& $\ 27.88$ & $\ 27.79$ & $\ 27.77$\\ \hline
		Self-Energy & $-35.88$ & $-35.88$ &$-35.15$& $-36.00$ & $-35.88$ & $-35.03$\\ \hline
			
	\end{tabular}
	\caption{Results of continuum extrapolation on gluonless ensembles for Fourier method with $\Lambda =3\, m_\mu$. The values are given in units of $10^{-11}$.
    }
	\label{tab::Tables_Fourier_gluonless_Results}

\end{table}
\begin{table}[ht]
    \centering

	\begin{tabular}{c c c c c c c}
		\hline
		 & XlYlZl & XcYlZl &XcYcZl & XlYlZc & XcYlZc & XcYcZc\\ \hline
		Total & $-7.41$ & $ -7.67$ & $-7.58$ & $-7.68$ & $ -7.84$ & $-7.61$\\ \hline
		2-Loop & $\ 28.48$ & $\ 28.34$ &$\ 28.29$& $\ 28.39$ & $\ 28.27$ & $\ 28.20$\\ \hline
		Self-Energy & $-35.88$ & $-36.01$ &$-35.87$& $-36.08$ & $-36.11$ & $-35.81$\\ \hline
			
	\end{tabular}
	\caption{Results of continuum extrapolation on gluonless ensembles for 5D propagator method with $\Lambda= 3\, m_\mu$. The values are given in units of $10^{-11}$.
    }
	\label{tab::Tables_5D_gluonless_Results}
\end{table}

\begin{table}[ht]
    \centering

	\begin{tabular}{c c c c c}
		\hline
		 & XlYlZl & XcYlZl & XlYlZc & XcYlZc \\ \hline
		2PS + Fourier & $-7.82$ & $ -8.18$ & $-8.08$ & $-8.28$ \\ \hline
		2PS + 5D propagator & $-7.29$ & $-7.70$ &$-7.57$& $-7.80$ \\ \hline
			
	\end{tabular}
	\caption{Results of continuum extrapolation on gluonless ensembles for calculating (4)a diagram via 2PS method and (4)b diagram via one of the other methods with $\Lambda= 3\, m_\mu$. The values are given in units of $10^{-11}$.
    }
	\label{tab::Tables_Hybrid_Extrapolation_Results}
\end{table}

\newpage
\bibliographystyle{JHEP}
\bibliography{references}
\end{document}